\documentclass{new_tlp}
\usepackage{mathptmx}

\renewcommand{\to}{\ensuremath{\rightarrow}}
\usepackage{microtype}

\usepackage{amsmath}

\usepackage{xspace}
\usepackage{color}
\usepackage{enumerate}
\usepackage{enumitem}
\usepackage{amssymb}
\usepackage{hyperref}
\usepackage{cleveref}
\usepackage{mathtools}
\usepackage{natbib}

\DeclareMathAlphabet{\mathcal}{OMS}{cmsy}{m}{n}

\renewcommand{\cite}[1]{\citep{#1}}

\usepackage[disable]{todonotes}   %

\definecolor{asparagus}{rgb}{0.53, 0.66, 0.42}

\usepackage{comment}

\newcommand{\dyna}{Dyna\xspace}

\newcommand{\defn}[1]{\textbf{#1}}
\newcommand{\func}[1]{{\tt \textbf{#1}}\xspace}

\usepackage[nottdefault]{sourcecodepro}

\newcommand{\rexpr}{{\def\ttdefault{SourceCodePro-TLF}%
    \text{{\tt \textbf{R}}-expr}}\xspace}
\newcommand{\rexprs}{{\def\ttdefault{SourceCodePro-TLF}%
\text{{\tt \textbf{R}}-exprs}}\xspace}

\newcommand{\todocolor}[1]{\color{#1}}
\renewcommand{\todocolor}[1]{}

\usepackage{cprotect}
\usepackage{wrapfig}
\usepackage{listings}

\lstdefinelanguage{dyna}{
  morekeywords={for,with\_key,arg},
  sensitive=true,
  morecomment=[l]{\%},
  morestring=[b]",
    literate=
    {:-}{{{\color{aggblue}:-}}}2
    {=}{{{\color{aggblue}=}}}1   
    {min=}{{{\color{aggblue}min=}}}4
    {max=}{{{\color{aggblue}max=}}}4
    {+=}{{{\color{aggblue}+=}}}2
    {|=}{{{\color{aggblue}|=}}}2
    {&=}{{{\color{aggblue}\&=}}}2
}

\lstnewenvironment{abstractex}[1][]
                  {%
                    \renewcommand{\ttdefault}{zi4}  
                    \setcounter{dynaGlobalLineNumber}{\the\value{lstnumber}}%
                          \lstset{
                      numbers=none,
                      language={},
                      mathescape=true,
                      xleftmargin={0cm},
                      frame=none,
                      #1
                    }%
                    \vspace{-2pt}
                  }{
                    \setcounter{lstnumber}{\the\value{dynaGlobalLineNumber}}%
                    \vspace{-2pt}
                  }

\lstnewenvironment{dynaex}[2][]
                  {%
                    \lstset{                    
                      language={dyna},
                      morekeywords={for,with\_key,arg},
                      classoffset=2,
                     morekeywords={A,B,C,D,E,F,G,H,I,J,K,L,M,N,O,P,Q,R,S,T,U,V,W,X,Y,Z,Xs,Ys,Zs,HX,HY,DX,DY,DZ,XX,YY,Property,#2},
                      keywordstyle=\color{vargreen},
                      #1
                    }
                    \vspace{-2pt}
                  }
                  {%
                    \vspace{-2pt}
                  }

\lstnewenvironment{termex}[2][]
                  {%
                    \renewcommand{\ttdefault}{zi4}  
                    \setcounter{dynaGlobalLineNumber}{\the\value{lstnumber}}%
                    \lstset{  
                      language={},
                      mathescape=true,
                      numbers=none,
                      classoffset=1,
                      morekeywords={when,if,where,and},
                      keywordstyle=\tt\bf,
                      xleftmargin={0cm},
                      frame=none,
                      classoffset=2,
                     morekeywords={G,H,P,A,B,C,D,E,F,I,J,K,O,T,U,V,W,X,Y,Z,Xs,Ys,Zs,HX,HY,DX,DY,DZ,Key,Val,Val1,Result,RSum,RMin,REq,Temp,T1,Tn,Xn,X1,Arg1,Arg2,XX,YY,Property},
                      keywordstyle=\color{vargreen},
                      classoffset=3,
                      morekeywords={R,S,Q,R2,R3,R0,R1#2},
                      keywordstyle=\color{termvar},
                      classoffset=4,
                      morekeywords={},
                      keywordstyle=\color{termground},
                      classoffset=5,
                      morekeywords={L,M,N},
                      keywordstyle=\color{termmul},
                      #1
                    }
                    \vspace{-2pt}
                  }
                  {%
                    \setcounter{lstnumber}{\the\value{dynaGlobalLineNumber}}%
                    \vspace{-2pt}
                  }

\newcommand{\rterm}[1]{%
\renewcommand{\ttdefault}{zi4}
\text{\lstinline[language={},  
                      mathescape=true,
                      numbers=none,
                      classoffset=1,
                      morekeywords={when,if,where,and},
                      keywordstyle=\tt\bf,
                      xleftmargin={0cm},
                      frame=none,
                      classoffset=2,
                     morekeywords={G,H,P,A,B,C,D,E,F,I,J,K,O,T,U,V,W,X,Y,Z,Xs,Ys,Zs,HX,HY,DX,DY,DZ,Key,Val,Val1,Result,RSum,RMin,REq,T1,Tn,Xn,X1,Temp,Arg1,Arg2},
                      keywordstyle=\color{vargreen},
                      classoffset=3,
                      morekeywords={R,S,Q,R2,R3,R0,R1},
                      keywordstyle=\color{termvar},
                      classoffset=4,
                      morekeywords={},
                      keywordstyle=\color{termground},
                      classoffset=5,
                      morekeywords={L,M,N},
                      keywordstyle=\color{termmul},
                      ]{#1}}\xspace%
}

\newcommand{\dynainline}[1]{%
\text{\lstinline[language={dyna},
                 morekeywords={for,with\_key,arg},
                 classoffset=2,
                 morekeywords={A,B,C,D,E,F,G,H,I,J,K,L,M,N,O,P,Q,R,S,T,U,V,W,X,Y,Z,Xs,Ys,Zs,HX,HY,DX,DY,DZ,Start,End,XX,YY,Property},
                 keywordstyle=\color{vargreen}
]{#1}}\xspace}

\definecolor{aggblue}{rgb}{0,0,.65}

\definecolor{vargreen}{rgb}{0,.50,.18}

\definecolor{termvar}{rgb}{0.0,.55,.55}  
\definecolor{termground}{rgb}{0.8, 0.5, 0.2}
\definecolor{termmul}{rgb}{0.03, 0.27, 0.49} 

\newcounter{dynaGlobalLineNumber}

\newcommand{\var}[1]{{\renewcommand{\ttdefault}{zi4}
\texttt{{\color{vargreen} #1}}}}

\newcommand{\vI}{\var{I}\xspace}
\newcommand{\vJ}{\var{J}\xspace}
\newcommand{\vK}{\var{K}\xspace}

\newcommand{\vR}{\var{R}\xspace}

\newcommand{\vT}{\var{T}\xspace}

\newcommand{\vX}{\var{X}\xspace}
\newcommand{\vXone}{\var{X}\mbox{$_1$}\xspace}
\newcommand{\vXtwo}{\var{X}\mbox{$_2$}\xspace}

\newcommand{\theMults}{\overline{\mathbb{N}}}

\usepackage{stmaryrd}  %
\newcommand{\sem}[1]{\left\llbracket #1 \right\rrbracket}

\definecolor{typeannotgray}{rgb}{0.4,.4,.4}

\usepackage[plain]{algorithm}
\usepackage{algorithmicx}
\usepackage[noend]{algpseudocode}

\MakeRobust{\Call}   %

\MakeRobust{\Call}   %
\newcommand{\rightcomment}[1]{\(\triangleright\) {\footnotesize\textit{#1}}}
\algrenewcommand{\algorithmiccomment}[1]{\hfill \rightcomment{#1}}  %
\algnewcommand{\LinesComment}[1]{\State \rightcomment{\parbox[t]{\linewidth-\leftmargin-\widthof{42:   }-\widthof{\(\triangleright\) }}{#1}}\smallskip}
\algnewcommand{\LineComment}[1]{\LinesComment{#1}}

\crefname{section}{\S}{\S\S}
\Crefname{section}{\S}{\S\S}
\crefformat{section}{#2\S#1#3}  %
\Crefformat{section}{#2\S#1#3}

\Crefname{appendix}{\S}{\S\S}
\crefname{appendix}{\S}{\S\S}
\crefformat{appendix}{#2\S#1#3}
\Crefformat{appendix}{#2\S#1#3}

\definecolor{darkmidnightblue}{rgb}{0.0, 0.2, 0.4}

\newcommand{\vars}{\mathrm{vars}}
\newcommand{\dom}{\mathrm{domain}}

\newcommand{\comm}[1]{\hfill\rm$\triangleright$\textit{#1}}

\newcounter{rtermrewrite}
\newcommand{\rto}{\ensuremath{\refstepcounter{rtermrewrite}{\todocolor{red}\xrightarrow{{\todocolor{blue!50}\footnotesize\thertermrewrite}}}}\xspace}
\newcommand{\rtoleftright}{\refstepcounter{rtermrewrite}{\todocolor{red}\ensuremath{\xleftrightarrow{{\todocolor{blue!50}\footnotesize\thertermrewrite}}}}\xspace}

\crefname{rtermrewrite}{rewrite rule}{rewrite rules}

\title[A Calculus for Bag Relations]{Evaluation of Logic Programs with Built-Ins and Aggregation: A Calculus for Bag Relations}

\author[M. Francis-Landau]
    {Matthew Francis-Landau and Tim~Vieira and Jason~Eisner \\
      Johns Hopkins University, Baltimore Maryland, USA \\
      \email{\{mfl,timv,jason\}@cs.jhu.edu}}

\jdate{June 2020}
\pubyear{2020}
\pagerange{\pageref{firstpage}--\pageref{lastpage}}
\doi{S1471068401001193}

\submitted{5 June 2020}
\revised{XXX}
\accepted{XXX}

\begin{document}

\label{firstpage}

\maketitle

\begin{abstract}

  We present a scheme for translating logic programs, which may use aggregation and arithmetic, into algebraic expressions that denote bag relations over ground terms of the Herbrand universe.  To evaluate queries against these relations,
  we develop an operational semantics based on term rewriting of the algebraic
  expressions.  This approach can exploit arithmetic identities and recovers a
  range of useful strategies, including lazy strategies that defer work until it becomes possible or necessary.\footnote{\label{footnote:pubinfo}\label{footnote:code} This manuscript is an expanded version of a paper that was presented at the 13th International Workshop on Rewriting Logic and Its Applications (WRLA 2020).  Code is available at
  \url{https://github.com/matthewfl/dyna-R}.}
\end{abstract}

  \begin{keywords}
    Logic Programming, Relational Algebra, Term Rewriting%
  \end{keywords}

\section{Introduction}\label{sec:intro}

We are interested in developing execution strategies for deductive databases
whose defining rules make use of aggregation, recursion, and arithmetic. Languages for specifying such deductive databases are expressive
enough that it can be challenging to answer queries against a given database.
Term rewriting systems are an attractive approach because they can start with
the query-database pair itself as an intensional description of the answer, and
then attempt to rearrange it into a more useful extensional description such as
a relational table.
We will give a notation for algebraically constructing potentially
infinite bag relations
from constants and built-in relations by unions, joins, projection, aggregation,
and recursion.  We show how programs in existing declarative languages such as
Datalog, pure Prolog, and Dyna can be converted into this notation.\looseness=-1

We will then sketch a term rewriting system for simplifying these algebraic expressions, which can be used to answer queries against a bag relation.  Term rewriting
naturally handles
materialization,            %
delayed constraints,        %
constraint propagation,     %
short-circuit evaluation,   %
lazy iteration,             %
and enumerative strategies  %
such as nested-iterator join.
  Our current implementation$^{\text{\ref{footnote:code}}}$
  is an interpreter that manages the application of these rewrites using a specific strategy outlined in \cref{sec:exec_with_r} (along with a memoization facility that is beyond the scope of this paper).  We are also exploring the use of machine learning to explore rewriting strategies \cite{mapl2017}.

\subsection{Approach}

\newcommand{\mydefs}[1]{{\color{black}#1}}

\newcommand{\theFunctors}[0]{\mydefs{\mathcal{F}}}
\newcommand{\theGround}[0]{\mydefs{\mathcal{G}}}
\newcommand{\theVars}[0]{\mydefs{\mathcal{V}}}
\newcommand{\theTerms}[0]{\mydefs{\mathcal{T}}}
\renewcommand{\theMults}[0]{\mydefs{\mathcal{M}}}   %
\newcommand{\theRexprs}[0]{\mydefs{\mathcal{R}}}
\newcommand{\rNull}[0]{\mydefs{\textsf{null}}}
\newcommand{\rTrue}[0]{\mydefs{\textsf{true}}}

Dyna~\cite{eisner-filardo-2011} is a generalization of
Datalog~\cite{datalog,datalog2} and pure Prolog~\cite{prolog,prolog2}.  Our
methods apply to all three of these logic programming languages.  Our methods could also be used on simpler languages like SQL that correspond to standard relational algebra, which we generalize.

We are given a Herbrand universe $\theGround$ of ground terms.  A Dyna program
serves to define a partial map from $\theGround$ to $\theGround$, which may be
regarded as a set of key-value pairs.  A Datalog or Prolog program is similar, but it
can map a key in $\theGround$ only to $\rTrue$, so it serves only to define the set of
keys.\looseness=-1

Given a program, a user may query the value of a specific key (ground term).
More generally, a user may use a non-ground term to query the values of all keys
that match it, so that the answer is itself a set of key-value pairs.

A set of key-value pairs---either a program or the answer to a query---is a
relation on two $\theGround$-valued variables.  Our method in this paper will be
to describe the desired relation algebraically, building up the description from
simpler relations using a relational algebra.  These simpler relations can be
over any number of variables; they may or may not have functional dependencies
as the final key-value relation does; and they may be bag relations, i.e., a
given tuple may appear in the relation more than once, or even infinitely many
times.  Given this description of the desired relation, we will use rewrite
rules to simplify it into a form that is more useful to the user.  Although we
use a \defn{term rewriting system}, we refer to the descriptions being rewritten
as \rexprs (relational expressions) to avoid confusion with the terms of the
object language, Dyna.

\subsection{The Dyna Language}

Datalog is a logical notation for defining finite database relations.  Using Prolog's pattern-matching notation, a Datalog program can concisely define each relation from other relations.  The notation makes it easy to express constants, don't-cares, union, selection, and join.  For example, a join of relations \dynainline{b} and \dynainline{c} is defined by the Datalog rule \dynainline{a(I,K) :- b(I,J), c(J,K)}, where the capitalized identifiers are variables.  The \dynainline{:-} operator is read as ``if,'' and the top-level comma is read as ``and.'' Thus, if \dynainline{b(2,8)} and \dynainline{c(8,5)} have already been established to be true, then this rule will establish that the ground term $\dynainline{a(2,5)} \in \theGround$ is true---that is, the defined relation \dynainline{a} contains the tuple $(2,5)$.

While Datalog permits relations to be defined recursively and sometimes provides access to particular infinite arithmetic relations, certain restrictions in Datalog ensure that all new relations defined by a user program remain finite and hence materializable.  \dyna drops these restrictions to allow the definition of infinite relations as in Prolog.

\dyna also provides an attractive notation for aggregation.  A \dyna rule is not a \defn{Horn clause} (as above) but a \defn{Horn equation} such as \rterm{a(I,K) += b(I,J) * c(J,K)}.  This rule can be read as implementing a generalized form of the matrix multiplication $(\forall i,j) a_{ik} = \sum_j b_{ij} \cdot c_{jk}$. It is ``generalized'' because \dynainline{I}, \dynainline{J}, \dynainline{K} range not over some finite set of integers, but over the entire Herbrand universe $\theGround$.  Their values may be any constant terms, including structured terms such as lists, perhaps representing real-world entities.  Nonetheless, for a given \dynainline{I} and \dynainline{K}, it may be that \dynainline{a(I,K)} has only finitely many summands, because summands without values are dropped, as we will see in a moment.

The difference from Horn clauses is that 
\dynainline{a(I,K) += b(I,J) * c(J,K)} is used not to establish that \dynainline{a(2,5)} is true, but that \dynainline{a(2,5)} has a particular \defn{value} $V$.  This may be regarded as saying that the defined relation \dynainline{a} contains the tuple $(2,5,V)$.  A ground term has at most one value, so this can be true for at most one $V$.  The \dynainline{+=} \defn{aggregation operator} indicates that this $V$ is obtained as $\sum_{J} \rterm{b(2,J)} \cdot \rterm{c(J,5)}$, where the summation is over all \dynainline{J} for which \dynainline{b(2,J)} and \dynainline{c(J,5)} both have values.  If there are no summands, then \dynainline{a(2,5)} has no value, i.e., \dynainline{a} contains no tuple of the form $(2,5,V)$.  Other rules may use other aggregation operators.  If \dynainline{a(2,5)} unifies with the heads of multiple rules, then its value is obtained by aggregating over the instantiated bodies contributed by all of those rules (all of which must have the same aggregation operator).

These and other features of \dyna make it possible to express many kinds of AI computations concisely and declaratively, as shown by \citet{eisner-filardo-2011}. Their examples include graph algorithms such as shortest path and edit distance, finite and infinite neural networks, non-monotonic reasoning, constraint satisfaction via backtracking search and constraint propagation, exact parsing with probabilistic grammars, approximate inference in graphical models using message passing or Monte Carlo methods, and Markov decision processes.  Each of their programs is essentially a system of Horn equations that must be solved to fixpoint.  The use of variables in the rules makes it possible to specify a large system of equations with just a few rules.  Fundamentally, the insight is that a computation can often be described using pattern-matching rules that describe the dependence of each named quantity on other named quantities.

Naive approaches to evaluating \dyna programs---the pure backward-chaining and forward-chaining strategies that are traditionally used in Prolog and Datalog respectively---often fail to terminate on real examples like the ones given by \citet{eisner-filardo-2011}.  This may occur either because of ``infinite depth'' (the chaining proceeds forever, continuing to visit new terms or cyclically revisit old ones) or because of ``infinite breadth'' (a single step of chaining touches infinitely many ground terms because a variable remains free).  

We therefore propose a term-rewriting approach that is capable of rearranging computations over infinite relations and which can eliminate parts of a computation that are guaranteed not to affect the result.  We will represent a Dyna program---or the intersection of a program with a query---using a finite algebraic object, namely an \rexpr, which defines a set of key-value pairs.  A step of execution then consists of rewriting this \rexpr in a semantics-preserving way.  We will define \rexprs and the rewrite rules in \cref{sec:syntax_semantics,sec:rewrite_rules}.

In many cases, it is possible to eventually simplify the \rexpr into a materialized finite relation.  Even when the \rexpr represents an infinite relation, it may still be possible to simplify it into a short expression that is readily interpretable by the human who issued the query (e.g., a materialized relation with don't-care variables, or a disjunction of conjunctions of arithmetic constraints).  
This ``best-effort'' behavior resembles that of a computer algebra system for simplifying mathematical expressions, or a constraint logic programming solver that may return answers with delayed constraints.

\subsection{Two \dyna Examples}\label{sec:dynaex}

To illustrate why \dyna execution is difficult, we discuss two examples.  The following program is formally quite similar to the previous matrix multiplication example.  It defines all-pairs shortest paths in a directed graph. Each graph edge is defined by an \dynainline{edge} term whose value is the length of that edge.  For any \dynainline{Start} and \dynainline{End} such that there exists a directed path from \dynainline{Start} to \dynainline{End}, the value of $\dynainline{path(Start,End)}$ is the total length (via \dynainline{+}) of the \emph{shortest} such path (thanks to the aggregation operator \dynainline{min=}).

\begin{dynaex}{Start,Mid,End}
path(Start,Start) min= 0.  @\hfill @% base case @\label{line:datalog_distance}@
path(Start,End) min= path(Start,Mid) + edge(Mid,End). @{\hfill\color{black!65} \% recursive case}\label{line:datalog_edge}@ @ @
edge("baltimore", "washington dc") = 38.  @\hfill @% example rules for edges @ @
edge("baltimore", "new york") = 195.  
@$\vdots$@  @\hfill @% many other edge rules omitted @ @
\end{dynaex}

This Dyna program is not a valid Datalog program, because the base case in line~\ref{line:datalog_distance} establishes an infinite set of facts.\footnote{The first version of Dyna \cite{eisner-goldlust-smith-2005} also forbade this program, so as to allow a Datalog-like execution strategy.}
This could be remedied by changing line~\ref{line:datalog_distance} to \dynainline{path(Start,Start) min= 0} \dynainline{for city(Start)}, thus restricting \dynainline{Start} to fall in a finite set of cities rather than ranging over all of $\theGround$.
However, the original program is meaningful mathematically, so we would still like to be able to \defn{query} the \defn{path} relation that it defines---i.e., obtain a compact description of the set of all ground terms that match a given pattern, along with their values.  (Terms with no value are not returned.)  For example, the query \dynainline{path(X,Y)} should return an infinite set; the query \dynainline{path("atlantis",Y)} should return a finite set.  Both of these sets contain \dynainline{path("atlantis","atlantis")}.)  

A Prolog-style solver does allow this program, and attempts to answer these queries by depth-first backward chaining that lazily instantiates variables with unification.  A query's results may include non-ground terms such as \dynainline{path(Start,Start)} that represent infinite sets.  Unfortunately, the Prolog strategy will recurse infinitely on line~\ref{line:datalog_edge}, even on the query \dynainline{path("atlantis",Y)}.  

Yet \dynainline{path("atlantis",Y)} is simply a single-source shortest path problem that \emph{could} be answered by \citeauthor{dijkstra-1959}'s \citeyear{dijkstra-1959} algorithm.  Similarly, the query \dynainline{path(X,Y)} \emph{could} be answered by running an all-pairs shortest path algorithm (e.g., Bellman-Ford) and augmenting the returned list of pairs with \dynainline{path(Start,Start)}, which represents an infinite set of pairs.  Our challenge is to construct a general Dyna engine that is capable of finding such algorithms.\footnote{Ideally, it should also be able to handle infinite graphs that are defined by \dynainline{edge} rules with variables, such as \dynainline{edge(N,N+1) = 1}.  Such a graph appears in the next example.}

Our second example is a Dyna program that defines a convolutional neural network over any two-dimensional grayscale image.  Larger images have more pixels; if there is a pixel at position \dynainline{(X,Y)}, its intensity is given as the value of \dynainline{pixel_brightness(X,Y)}.  
Lines~\ref{line:conv_start}--\ref{line:conv_out} define a standard feed-forward network architecture in which each unit is named by a term in $\theGround$, such as \dynainline{input(12,34)}, \dynainline{hidden(14,31)}, or \dynainline{output(kitten)}.  The activation \dynainline{out(J)} of unit \dynainline{J} is normally computed as a sigmoided linear combination of the activations of other units \dynainline{I}.  The specific topology is given by the \dynainline{edge(I,J)} items that connect \dynainline{I} to \dynainline{J}, which in this case are defined at lines~\ref{line:conv_edge}--\ref{line:conv_output}.  Specifically, \dynainline{hidden(XX,YY)} is activated by the
$9 \times 9$ square of neurons centered at \dynainline{input(XX,YY)}.  Then for each image property, \dynainline{Property}, all of the hidden units are pooled to activate an output unit, \dynainline{output(Property)}, whose activation represents the degree to which the image is predicted to have that property.

\begin{dynaex}{}
@$\sigma$@(X) = 1/(1+exp(-X)).     @\hfill @ % define sigmoid function for all X@\label{line:conv_start}@
in(J) += out(I) * edge(I,J).     @\hfill @ % vector-matrix product@\label{line:conv_in}@
out(J) += @$\sigma$@(in(J)).        @\hfill @ % activation of node J@\label{line:conv_out}@
out(input(X,Y)) += pixel_brightness(X,Y) @\hfill @ % extra external activation for input nodes
loss += (out(J) - target(J))**2. @\hfill @ % L2 loss of the predictions
edge(input(X,Y),hidden(X+DX,Y+DY)) = weight_conv(DX,DY). @\label{line:conv_edge}\hfill@ % layer 1
edge(hidden(XX,YY),output(Property)) = weight_output(Property).@\label{line:conv_output}\hfill@ % layer 2
weight_conv(DX,DY) := random(*,-1,1) for DX:-4..4, DY:-4..4. @\hfill @ % init with random@\label{line:conv_weight}@
weight_output(Property) := random(*,-1,1).@\label{line:output_weight}@
\end{dynaex}

For any given finite image, the program effectively defines a neural network with finitely many edges.  Nonetheless, it is difficult to solve this system of equations using standard strategies.  The difficulty arises from line~\ref{line:conv_edge}.  Using a forward chaining strategy (as is typical for Datalog), the existence of a value for \dynainline{weight_conv(2,-3)} would forward-chain through line~\ref{line:conv_edge} to establish values (i.e., weights) for infinitely many edges of the form \dynainline{edge(input(X,Y),hidden(XX,YY))} where $\dynainline{XX}=\dynainline{X}+2$ and $\dynainline{YY}=\dynainline{Y}-3$ ---even though only finitely many of these edges will turn out to be used when analyzing the finite image.  Conversely, using backward chaining (as in prolog) to answer the (useful) query \dynainline{out(output(kitten))} would lead to a subgoal query \dynainline{edge(I,hidden(XX,YY))} with free variables \dynainline{XX} and \dynainline{YY}.  That query must return the entire infinite set of input-to-hidden edges---even though for a given finite image, only finitely many of these edges will touch input units that actually have values.\footnote{As another example, backward-chaining a query \dynainline{edge(input(123,45),J))} to find the outgoing edges from a given input unit would lead to querying infinitely many weights of the form \dynainline{weight_conv(DX,DY)} where $\dynainline{DX}=\dynainline{XX}-123$ and
$\dynainline{DY}=\dynainline{YY}-45$---even though only finitely many of these weights will turn out to have values.}  These infinite sets in all of these examples cannot be represented by simple non-ground terms, because the variable arguments to \dynainline{input}, \dynainline{hidden}, and \dynainline{weight_conv} in line~\ref{line:conv_edge} are related by arithmetic rather than simply by unification.  However, our \rexpr formalism makes it possible to keep track of these arithmetic constraints among the variables.  \cref{sec:neural_ex} describes how our approach handles this example program.

\section{Syntax and Semantics of \protect\rexprs} \label{sec:syntax_semantics}

\newcommand{\rid}[1]{\mydefs{\ensuremath{\textsf{id}_{\tt #1}}}}
\renewcommand{\func}[1]{\texttt{#1}}
\newcommand{\rProj}[0]{\func{proj}}
\newcommand{\rPlus}[0]{\func{plus}}
\newcommand{\rExist}[0]{\func{exists}}
\newcommand{\rFailfast}[0]{\func{failfast}}
\newcommand{\rSum}[0]{\func{sum}}

\newcommand{\env}[0]{\mydefs{E}}
\newcommand{\rSem}[2]{\mydefs{\sem{#1}_{#2}}}  %
\newcommand{\rSemE}[1]{\rSem{#1}{\env}}  %

Let $\theGround$ be the Herbrand universe of \defn{ground terms} built from a
given set $\theFunctors$ of ranked functors.  We treat constants (including
numeric constants) as 0-ary functors.  Let
$\theMults = \mathbb{N} \cup \{\infty\}$ be the set of \defn{multiplicities}.  A
simple definition of a \defn{bag relation} \cite{Green2009bag-semantics} would
be a map $\theGround^n \mapsto \theMults$ for some $n$.  Such a map would
assign a multiplicity to each possible \emph{ordered $n$-tuple} of ground terms.
However, we will use names rather than positions to distinguish the roles of the
$n$ objects being related: in our scheme, the $n$ tuple elements will be named
by variables.

Let $\theVars$ be an infinite set of distinguished \defn{variables}.  A
\defn{named tuple} $\env$ is a function mapping some of these variables to
ground terms.  For any $\mathcal{U} \subseteq \theVars$, we can write
$\theGround^{\mathcal{U}}$ for the named tuples
$\env: \mathcal{U} \mapsto \theGround$ with domain $\mathcal{U}$.
These are just the possible \defn{$\mathcal{U}$-tuples}: that is, tuples over
$\theGround$ whose elements are named by $\mathcal{U}$.  This set
$\theGround^{\mathcal{U}}$ of named $\mathcal{U}$-tuples replaces the set
$\theGround^n$ of ordered $n$-tuples from above.\looseness=-1

Below, we will inductively define the set $\theRexprs$ of \rexprs.  The reader
may turn ahead to later sections to see some examples, culminating in \cref{sec:trans}, which gives a scheme to translate a Dyna program into an \rexpr.

Each \rexpr \rterm{R}
has a finite set of \defn{free variables} $\vars(\rterm{R}) \subseteq \theVars$,
namely the variables that appear in \rterm{R} in positions where they are not
bound by an operator such as \rterm{proj} or \rterm{sum}.  The idea is for
\rterm{R} to specify a bag relation over domain $\theGround$, with columns named
by $\vars(\rterm{R})$.\looseness=-1

The \defn{denotation function} $\rSemE{\cdot}$ interprets \rexprs in the
\defn{environment} given by the named tuple $\env$.  It defines a multiplicity
$\rSemE{\rterm{R}}$ for any \rexpr \rterm{R} whose
$\vars(\rterm{R}) \subseteq \dom(\env)$.

If $\mathcal{U} \supseteq \vars(\rterm{R})$, we can dually regard \rterm{R} as
inducing the map $\theGround^{\mathcal{U}} \to \theMults$ defined by
$\env \mapsto \rSemE{\rterm{R}}$.  In other words, given $\mathcal{U}$,
\rterm{R} specifies a bag relation whose column names are $\mathcal{U}$.  This
is true for \emph{any} $\mathcal{U} \supseteq \vars(\rterm{R})$, but the
relation constrains only the columns $\vars(\rterm{R})$.  The other columns can
take any values in $\theGround$.  A tuple's multiplicity never depends on its
values in those other columns, since our definition of $\rSemE{\cdot}$ will
ensure that $\rSemE{\rterm{R}}$ depends only on the restriction of $\env$ to
$\vars(\rterm{R})$.

We say that \rterm{T} is a \defn{term} if $\rterm{T} \in \theVars$ or
$\rterm{T} = f(\rterm{T}_1,\ldots,\rterm{T}_n)$ where $f \in \theFunctors$ has
rank $n$ and $\rterm{T}_1,\ldots,\rterm{T}_n$ are also terms.  Terms typically
appear in the object language (e.g., Dyna) as well as in our meta-language
(\rexprs).  Let $\theTerms \supseteq \theGround$ be the set of terms.  Let
$\vars(\rterm{T})$ be the set of vars appearing in \rterm{T}, and extend $E$ in
the natural way over terms \rterm{T} for which
$\vars(\rterm{T}) \subseteq \dom(\env)$:
$\env(f(\rterm{T}_1,\ldots,\rterm{T}_n)) =
f(\env(\rterm{T}_1),\ldots,\env(\rterm{T}_n))$.

We now define $\rSemE{\rterm{R}}$ for each type of \rexpr \rterm{R}, thus also
presenting the different types of \rexprs in $\theRexprs$.  First, we have
\defn{equality constraints} between non-ground terms, which are true in an
environment that grounds those terms to be equal.  True is represented by
multiplicity 1, and false by multiplicity 0.
\begin{enumerate}
    \item $\rSemE{ \rterm{T=U} }  = \textbf{ if } \env(\rterm{T})=\env(\rterm{U}) \textbf{ then } 1 \textbf{ else } 0$,
    \qquad where $\rterm{T},\rterm{U}\in\theTerms$
\end{enumerate}
Notice that we did not write $\rSemE{\rterm{T}}=\rSemE{\rterm{U}}$ but $\env(\rterm{T})=\env(\rterm{U})$---since our denotation function $\rSemE{\cdot}$ maps \rexprs to multiplicities, but $E(\cdot)$ maps non-ground terms in $\theTerms$ to ground terms in $\theGround$.

We also have \defn{built-in constraints}, such as
\begin{enumerate}[resume]
\item\label{item:builtin}$\rSemE{ \rterm{plus(I,J,K)} } = \textbf{ if } \env(\rterm{I}) + \env(\rterm{J}) = \env(\rterm{K}) \textbf{ then } 1 \textbf{ else } 0$,
\hfill where $\rterm{I},\rterm{J},\rterm{K}\in\theTerms$
\end{enumerate}
The above \rexprs are said to be \defn{constraints} because they always have
multiplicity 1 or 0 in any environment.  Taking the \defn{union} of \rexprs via
\rterm{+} may yield larger multiplicities:
\begin{enumerate}[resume]
\item $\rSemE{ \rterm{R+S} } = \rSemE{ \rterm{R} } + \rSemE{ \rterm{S} }$,
\qquad where $\rterm{R}, \rterm{S} \in \theRexprs$
\end{enumerate}
The \rexpr \rterm{0} denotes the empty bag relation, and more generally,
$\rterm{M}\in\theMults$ denotes the bag relation that contains \rterm{M} copies
of every $\mathcal{U}$-tuple:
\begin{enumerate}[resume]
\item $\rSemE{ \rterm{M} } = \rterm{M}$,
\qquad where $\rterm{M} \in \theMults$
\end{enumerate}
To \defn{intersect} two bag relations, we must use multiplication \rterm{*} to
combine their multiplicities \cite{Green2009bag-semantics}:
\begin{enumerate}[resume]
\item $\rSemE{ \rterm{R*S} } = \rSemE{ \rterm{R} } \cdot \rSemE{ \rterm{S} }$,
\qquad where $\rterm{R}, \rterm{S} \in \theRexprs$
\end{enumerate}
Here we regard both \rterm{R} and \rterm{S} as bag relations over columns
$\mathcal{U} \supseteq \vars(\rterm{R}) \cup \vars(\rterm{S}) =
\vars(\rterm{R*S})=\vars(\rterm{R+S})$.  The names \defn{intersection},
\defn{join} (or \defn{equijoin}), and \defn{Cartesian product} are
conventionally used for the cases of \rterm{R*S} where (respectively)
$\vars(\rterm{R})=\vars(\rterm{S})$, $|\vars(\rterm{R})\cap\vars(\rterm{S})|=1$,
and $|\vars(\rterm{R})\cap\vars(\rterm{S})|=0$.  As a special case of Cartesian
product, notice that \rterm{R*3} denotes the same bag relation as \rterm{R+R+R}.

We next define \defn{projection}, which removes a named column (variable) from a
bag relation, summing the multiplicities of rows (tuples) that have thus become
equal.  When we translate a logic program into an \rexpr (\cref{sec:trans}), we
will generally apply projection operators to each rule to eliminate that rule's
local variables.
\begin{enumerate}[resume]
\item\label{item:proj} $\rSemE{ \rterm{proj(X,R)} } = \sum_{x \in \theGround} \rSem{ \rterm{R} }{ \env[\rterm{X}=x] }$
\\ where $\rterm{X} \in \theVars, \rterm{R}\in \theRexprs$, and where $\env[\rterm{X}=x]$ means a version of $\env$ that has been modified to put $\env(\rterm{X})=x$ (adding \rterm{X} to its domain if it is not already there)
\end{enumerate}

Projection collapses each group of rows that are identical except for their
value of \rterm{X}.  \defn{Summation} does the same, but instead of adding up
the multiplicities of the rows in each possibly empty group, it adds up their
\rterm{X} values to get a \rterm{Y} value for the new row, which has
multiplicity 1.  Thus, it removes column \rterm{X} but introduces a new column
\rterm{Y}.  The summation operator is defined as follows (note that
$\rterm{sum} \notin \theFunctors$):
\begin{enumerate}[resume]
\item\label{item:aggregation} $\rSemE{ \rterm{A=sum(X,R)} } = \textbf{ if } \env(\rterm{A})=\sum_{x \in \theGround} x * \rSem{ \rterm{R} }{\env[\rterm{X}=x]} \textbf{ then } 1 \textbf{ else } 0$
\\
where $\rterm{A},\rterm{X} \in \theVars, \rterm{R} \in \theRexprs$.  The $*$ in the summand means that we sum up $\rSem{ \rterm{R} }{\env[\rterm{X}=x]}$ copies of the value $x$ (possibly $\infty$ copies).
If there are no summands, the result of the $\sum \cdots$ is defined to be the identity element \rid{sum}.
\end{enumerate}
Notice that an \rexpr of this form is a constraint.  \rterm{sum} is just one
type of \defn{aggregation operator}, based on the binary $+$ operation and its
identity element \rid{sum}$=0$. In exactly the same way, one may define other
aggregation operators such as \rterm{min}, based on the binary $\min$ operation
and its identity element \rid{min}$=\infty$.  Variants of these will be used in
\cref{sec:trans} to implement the aggregations in \rterm{+=} and \rterm{min=}
rules like those in \cref{sec:dynaex}.

In projections \rterm{proj(X,R)} and aggregations such as \rterm{sum(X,R)} and
\rterm{min(X,R)}, we say that occurrences of \rterm{X} within \rterm{R} are
\defn{bound} by the projection or aggregation operator\footnote{But notice that since
  \rterm{sum(X,R)} sums \rterm{X} values, it does not correspond to $\sum_{\rterm{X}} \cdots$ but rather
  to $\sum_{\text{row} \in \rterm{R}} \text{row}[\rterm{X}]$.} (unless they are bound by another operator within \rterm{R}).  Thus they are
not in the $\vars$ of the resulting \rexpr.  However, the most basic aggregation
operator does not need to bind a variable:
\begin{enumerate}[resume]
\item $\rSemE{ \rterm{M=count(R)} } = \textbf{ if } \env(\rterm{M})=\rSemE{ \rterm{R} } \textbf{ then } 1 \textbf{ else } 0$
\end{enumerate}
In effect, \rterm{M=count(R)} is a version of \rterm{R} that changes every
tuple's multiplicity to 1 but records its original multiplicity in a new column
\rterm{M}.  It is equivalent to \rterm{M=sum(N,(N=1)*R)} (where
$\rterm{N}\notin\vars(\rterm{R})$), but we define it separately here so that it
can later serve as an intermediate form in the operational semantics.

Finally, it is convenient to augment the built-in constraint types with
\defn{user-defined} relation types.  Choose a new functor of rank $n$ that is
$\notin \theFunctors$, such as $\rterm{f}$, and choose some \rexpr
$\rterm{R}_{\rterm{f}}$ with
$\vars(\rterm{R}_{\rterm{f}}) \subseteq \{\rterm{X}_1,\ldots,\rterm{X}_n\}$
(which are $n$ distinct variables) to serve as the \defn{definition} (macro
expansion) of \rterm{f}.  Now define
\begin{enumerate}[resume]
\item\label{item:userconstraint} $\rSemE{ \rterm{f(T}_1,\ldots,\rterm{T}_n\rterm{)} } = \rSemE{ \rterm{R}_{\rterm{f}}\{\rterm{X}_1\mapsto \rterm{T}_1,\ldots,\rterm{X}_n\mapsto \rterm{T}_n\} }$
\quad where $\rterm{T}_1,\ldots,\rterm{T}_n \in \theTerms \cup \theRexprs$
\\ The $\{\mapsto\}$ notation denotes substitution for variables, where bound variables of $\rterm{R}_{\rterm{f}}$ are renamed to avoid capturing free variables of the $\rterm{T}_i$.
\end{enumerate}
With user-defined relation types, it is possible for a user to write \rexprs
that are circularly defined in terms of themselves or one another (similarly to
a \texttt{let rec} construction in functional languages).  Indeed, a Dyna
program normally does this.  In this case, the definition of $\rSemE{\cdot}$ is
no longer a well-founded inductive definition.  Nonetheless, we can still
interpret the $\rSemE{\cdots} = \cdots$ equations in the numbered points above
as \emph{constraints} on $\rSemE{\cdot}$, and attempt to solve for a denotation
function $\rSemE{\cdot}$ that satisfies these constraints
\cite{eisner-filardo-2011}.
Some circularly defined \rexprs may be constructed so as to have unique
solutions, but this is not the case in general.\looseness=-1

\section{Rewrite Rules}  \label{sec:rewrite_rules}

Where the previous section gave a denotational semantics for \rexprs, we now
sketch an operational semantics.  The basic idea is that we can use rewrite
rules to simplify an \rexpr until it is either a finite materialized
relation---a list of tuples---or has some other convenient form. All of our
rewrite rules are semantics-preserving, but some may be more helpful than
others.

For some \rexprs that involve infinite bag relations, there may be no
way to eliminate all built-in constraints or aggregation operations.  The
reduced form then includes delayed constraints
(just as in constraint logic programming), or delayed aggregators.  Even so, conjoining
this reduced form with
a query can permit further simplification, so some queries against the \rexpr may still
yield simple answers.

\subsection{Finite Materialized Relations} \label{sec:finite_materialized}

We may express the finite bag relation shown at left by a simple
\defn{sum-of-products} shown at the
right.  In this example, the ground values being related are integers.
\begin{align}
g =
\left[
\begin{matrix}
\vXone & \vXtwo \\
\hline
1 & 1 \\
2 & 6 \\
2 & 7 \\
2 & 7 \\
5 & 7
\end{matrix}
\right]
&
\quad
\xRightarrow{\text{to \rexpr}}
\quad
&
\rterm{R}_{\rterm{g}} =
\begin{matrix}
&( & (\vXone=1) &*& (\vXtwo=1) & \\
&+ & (\vXone=2) &*& (\vXtwo=6) & \\
&+ & (\vXone=2) &*& (\vXtwo=7) & \\
&+ & (\vXone=2) &*& (\vXtwo=7) & \\
&+ & (\vXone=5) &*& (\vXtwo=7) & )\\
\end{matrix}\label{eq:ex:g-def}
\end{align}

\noindent We see that each individual row of the table (tuple) translates into a
product ($*$) of several ($\var{Variable}=\texttt{value}$) expressions, where
the variable is the column's name and the value is the cell in the table.  To
encode multiple rows, the \rexpr simply adds the \rexprs for the individual
rows.  When evaluated in a given environment, the \rexpr is a sum of products of
multiplicities.  But abstracting over possible environments, it represents a
union of Cartesian products of 1-tuples, yielding a bag relation.

We may use this \rexpr as the basis of a new user-defined \rexpr type \rterm{g}
(case~\ref{item:userconstraint} of \cref{sec:syntax_semantics}) by taking its
definition $\rterm{R}_{\rterm{g}}$ to be this \rexpr.  Our \rexprs can now
include constraints such as \rterm{g(J,K)} or \rterm{g(J,7)}.  When adding a new
case in the denotational semantics in this way, we always match it in the
operational semantics by introducing a corresponding rewrite rule
$\rterm{g(X}_1,\rterm{X}_2\rterm{)} \to \rterm{R}_{\rterm{g}}$.

A sum-of-products \rexpr simply enumerates the tuples of a bag relation,
precisely as a boolean expression in disjunctive normal form (that is, an
``or-of-ands'' expression) enumerates the satisfying assignments.  Just as in
the boolean case, a disjunct does not have to constrain every variable: it may
have ``don't care'' elements.  For example, \rterm{(J=1)*(K=1) + (J=2)}
describes an infinite relation because the second disjunct \rterm{J=2} is true
for any value of \rterm{K}.

\subsection{Equality Constraints and Multiplicity Arithmetic} \label{sec:equality_prop}\label{sec:mult_arith}

We may wish to query whether the relation $g$ in the above section relates $2$
to $7$.  Indeed it does---and twice.  We may discover this by considering
\rterm{g(2,7)}, which rewrites immediately via substitution to an \rexpr that
has no variables at all (\rterm{(2=1)*(7=1)+}$\cdots$), and finding that this
further reduces to the multiplicity \rterm{2}.

How does this reduction work?  First, we need to know how to evaluate the
equality constraints: we need to rewrite $\rterm{2=1} \to \rterm{0}$ but
$\rterm{2=2} \to \rterm{1}$.  The following structural unification rules provide the necessary rewrites:

\begin{termex}{}
($f$(U$_1$,$\ldots$,U$_n$)=$g$(V$_1$,$\ldots$,V$_m$)) $\rto\label{rr:struct_unify1}$ 0 if $f, g \in \theFunctors$ and $f \ne g$  @\comm{functor clash (note that $n\ne m \Rightarrow f\ne g$)}@
($f$(U$_1$,$\ldots$,V$_n$)=$f$(V$_1$,$\ldots$,V$_n$)) $\rto\label{rr:struct_unify2}$ (U$_1$=V$_1$) * $\cdots$ * (U$_n$=V$_n$) if $f \in \theFunctors$ and $n \geq 0$
(T=X) $\rto$ (X=T) if X$\,\in \theVars$ and T$\,\in \theTerms$ @\comm{put var on left to match rules below}@
(X=X) $\rto$ 1 if X$\,\in \theVars$   @\comm{because true for every value of $\vX$}@
(X=T) $\rto\label{rr:occurs_check}$ 0 if X$\,\in \theVars$ and T$\,\in \theTerms$ and X$\,\in \vars(\vT)$ @\comm{not true for any $\vX$ (occurs check)}@
\end{termex}
We now have an arithmetic expression \rterm{0*0+1*0+1*1+1*1+0*1}, which we can reduce to the multiplicity
\rterm{2} via rewrites that implement basic arithmetic on multiplicities
$\theMults$:%
\begin{termex}{}
M + N $\rto$ L if M,N $\in \theMults$ and M$+$N$=$L;   M * N $\rto\label{rr:multiply_mult}$ L if M,N $\in \theMults$ and M$*$N$=$L
\end{termex}

Above, we relied on the definition of the new relation type \rterm{g}, which
allowed us to request a specialization of $\rterm{R}_{\rterm{g}}$.  Do we need
to make such a definition in order to query a given bag relation?  No: we may do
so by conjoining the \rexpr with additional constraints.  For example, to get
the multiplicity of the pair $(2,7)$ in $\rterm{R}_{\rterm{g}}$, we may write
$\rterm{R}_{\rterm{g}}\rterm{*(X}_1\rterm{=2)*(X}_2\rterm{=7)}$.  This filters
the original relation to just the pairs that match $(2,7)$, and simplifies to
$\rterm{2*(X}_1\rterm{=2)*(X}_2\rterm{=7)}$.  To accomplish this simplification,
we need to use the following crucial rewrite rule:
\begin{termex}{}
(X=T)*R $\rto\label{rr:equality_prop}$ (X=T)*R$\{\texttt{X} \mapsto \texttt{T}\}$ if X $\in \theVars$ and T $\in \theTerms$ and X$\,\in \vars(\vR)$ @\comm{equality propagation}@
\end{termex}
As a more interesting example, the reader may consider simplifying the query
$\rterm{R}_{\rterm{g}}\rterm{*lessthan(X}_2\rterm{,7)}$, which uses a built-in
inequality constraint (see \cref{sec:built-ins}).

\subsection{Joining Relations} \label{sec:joining_relations}

Analogous to \cref{eq:ex:g-def}, we define a second tabular relation $f$ with a
rewrite rule $\rterm{f(X}_1,\rterm{X}_2\rterm{)} \to \rterm{R}_{\rterm{f}}$.

\vspace{-1\baselineskip}
\begin{align}
\qquad\quad
f =
\left[
\begin{matrix}
\vX_1 & \vX_2 \\
\hline
1 & 2 \\
3 & 4
\end{matrix}
\right]
&
\quad
\xRightarrow{\text{to \rexpr}}
\hspace{-1cm}
&
\rterm{R}_{\rterm{f}} =
\begin{matrix}
  & (\vXone=1) &*& (\vXtwo=2) \\
+ & (\vXone=3) &*& (\vXtwo=4)
\end{matrix}\label{eq:ex:f-def}
\end{align}
We can now consider simplifying the \rexpr\ \rterm{f(I,J) * g(J,K)}, which the
reader may recognize as an equijoin on $f$'s second column and $g$'s first
column.\footnote{This notation may be familiar from Datalog, except that we are
  writing the conjunction operation as \rterm{*} rather than with a comma, to
  emphasize the fact that we are multiplying multiplicities rather than merely
  conjoining booleans.}  Notice that the \rexpr has \emph{renamed} these columns
to its own free variables ($\vI$, $\vJ$, $\vK$).  Reusing the variable \vJ in
each factor is what gives rise to the join on the relevant column.  (Compare
\rterm{f(I,J) * g(J',K)}, which does not share the variable and so gives the
Cartesian product of the $f$ and $g$ relations.)

We can ``materialize'' the equijoin by reducing it to a sum-of-products form as before, if we wish:
\ \ \rterm{(I=1)*(J=2)*(K=6) + 2*(I=1)*(J=2)*(K=7)}.

To carry out such simplifications, we use the fact that multiplicities form a
commutative semiring under \rterm{+} and \rterm{*}.  Since any \rexpr evaluates
to a multiplicity, these rewrites can be used to rearrange unions and joins of
\rexprs $\rterm{Q}, \rterm{R}, \rterm{S} \in \theRexprs$:
\begin{termex}{}
1 * R $\rto\label{rr:one_mult}$ R @\comm{multiplicative identity}@
0 * R $\rto\label{rr:zero_mult}$ 0 @\comm{multiplicative annihilation}@
0 + R $\rto\label{rr:add_identity}$ R @\comm{additive identity}@
$\infty$ * R $\rto\label{rr:infity_mult}$ $\infty$ if R $\in \theMults$ and R $>$ 0 @\comm{absorbing element}@
$\infty$ + R $\rto\label{rr:infity_add}$ $\infty$ @\comm{absorbing element}@
R + S $\rtoleftright$ S + R;                R * S $\rtoleftright\label{rr:commutativity_times}$ S * R  @\comm{commutativity}@
Q + (R + S) $\rtoleftright$ (Q + R) + S;    Q * (R * S) $\rtoleftright$ (Q * R) * S    @\comm{associativity}@
Q * (R + S) $\rtoleftright\label{rr:distributivity}$ Q * R + Q * S  @\comm{distributivity}@
R * M $\rto$ R + (R * N) if M,N $\in \theMults$ and (1+N=M) @\comm{implicitly does \texttt{M} $\to$ \texttt{1+N}}@
R $\rtoleftright\label{rr:constraint-square}$ R * R if R @\textrm{is a constraint}@ @\comm{as defined in \cref{sec:syntax_semantics}}@
\end{termex}

We can apply some of these rules to simplify our example as follows:
\begin{termex}{}
f(I,J)*g(J,K)
@$\to$@ ((I=1)*(J=2)+(I=3)*(J=4)) * g(J,K)       @\comm{\cref{eq:ex:f-def}}@
@$\to$@ (I=1)*(J=2)*g(J,K) + (I=3)*(J=4)*g(J,K) @\comm{distributivity}@
@$\to$@ (I=1)*(J=2)*g(2,K) + (I=3)*(J=4)*g(4,K) @\comm{equality propagation}@
@$\to$@ (I=1)*(J=2)*( (K=6)+(K=7)*2 ) + (I=3)*(J=4)*0 @\comm{via \cref{eq:ex:g-def}}@
@$\to$@ (I=1)*(J=2)*( (K=6)+(K=7)*2 ) @\comm{annihilation}@
@$\to$@ (I=1)*(J=2)*(K=6) + (I=1)*(J=2)*(K=7)*2 @\comm{distributivity}@
\end{termex}

\noindent Notice that the factored intermediate form \rterm{(I=1)*(J=2)*(
  (K=6)*1 + (K=7)*2 )} is more compact than the final sum of products, and may
be preferable in some settings.  In fact, it is an example of a \defn{trie}
representation of a bag relation.  Like the root node of a trie, the expression
partitions the bag of (\rterm{I},\rterm{J},\rterm{K}) tuples into disjuncts
according to the value of \rterm{I}.  Each possible value of \rterm{I} (in this
case only \rterm{I=1}) is multiplied by a trie representation of the bag of
(\rterm{J},\rterm{K}) tuples that can co-occur with this \rterm{I}.  That
representation is a sum over possible values of \rterm{J} (in this case only
\rterm{J=2}), which recurses again to a sum over possible values of \rterm{K}
(\rterm{K=6} and \rterm{K=7}). Finally, the multiplicities \rterm{1} and
\rterm{2} are found at the leaves.

A trie-shaped \rexpr generally has a smaller branching factor than a
sum-of-products \rexpr.  As a result, it is comparatively fast to query it for
all tuples that restricts a prefix such as \rterm{I} or \rterm{(I,J)},
by narrowing down to the matching summand(s) at each node.
For example, multiplying our example trie by the query \rterm{I=5} gives an
\rexpr that can be immediately simplified to \rterm{0}, as the single disjunct
(for \rterm{I=1}) does not match.\looseness=-1

That example query also provides an occasion for a larger point.  This trie
simplification has the form \rterm{(I=5)*(I=1)*R}, an expression that in general
may be simplified to \rterm{0} on the basis of the first two factors, without
spending any work on simplifying the possibly large expression \rterm{R}.  This
is an example of \defn{short-circuiting} evaluation---the same logic that allows
a SAT solver or Prolog solver to backtrack immediately upon detecting a
contradiction.

\subsection{Rewrite Rules for Built-In Constraints}
\label{sec:built-ins}

Built-in constraints are an important ingredient in constructing infinite
relations.  While they are not the only method,\footnote{Others are structural
  equality constraints and recursive user-defined constraints.} they have the
advantage that libraries of built-in constraints such as \rterm{plus(I,J,K)}
(case~\ref{item:builtin} of \cref{sec:syntax_semantics}) usually come with
rewrite rules for reasoning about these constraints \cite{chr}.  Some of the
rewrite rules invoke opaque procedural code.

Recall that the arguments to a \rterm{plus} constraint are terms, typically
either variables or numeric constants.  Not all \rterm{plus} constraints can be
rewritten, but a library should provide at least the cases:\looseness=-1

\newcommand{\jasonVar}[1]{{\color{termground}\texttt{#1}}}
\newcommand{\rI}{\jasonVar{I}}
\newcommand{\rJ}{\jasonVar{J}}
\newcommand{\rK}{\jasonVar{K}}
\newcommand{\rR}{{\color{termvar}\texttt{R}}}
\newcommand{\rS}{{\color{termvar}\texttt{S}}}

\begin{termex}{}
plus($\rI$,$\rJ$,$\rK$) $\rto\label{rr:plus1}$ $\underbrace{\mathbb{I}(\rI+\rJ=\rK)}_{\in \{0,1\}}$ if $\rI,\rJ,\rK \in \mathbb{R}$
plus($\rI$,$\rJ$,X) $\rto\label{rr:plus2}$ (X=$\rI+\rJ$) if $\rI,\rJ \in \mathbb{R}$ and X $\in \theVars$
plus($\rI$,X,$\rK$) $\rto\label{rr:plus3}$ (X=$\rK-\rI$) if $\rI,\rK \in \mathbb{R}$ and X $\in \theVars$
plus(X,$\rJ$,$\rK$) $\rto\label{rr:plus4}$ (X=$\underbrace{\rK-\rJ}_{\in\mathbb{R}}$) if $\rJ,\rK \in \mathbb{R}$ and X $\in \theVars$
\end{termex}

\newcommand{\rRule}[2]{\rterm{#1}$\rto$ \rterm{#2}}

The \rexpr $\rterm{R}=\rterm{ proj(J,plus(I,3,J)*plus(J,4,K))}$ represents the
infinite set of $(\rterm{I},\rterm{K})$ pairs such that
$\rterm{K}=(\rterm{I}+3)+4$ arithmetically.  (The intermediate temporary
variable \rterm{J} is projected out.)  The rewrite rules already presented (plus
a rewrite rule from \cref{sec:proj} below to eliminate \rterm{proj}) suffice to
obtain a satisfactory answer to the query \rterm{I=2} or \rterm{K=9}, by
reducing either \rterm{(I=2)*R} or \rterm{R*(K=9)} to \rterm{(I=2)*(K=9)}.

On the other hand, if we wish to reduce \rterm{R} itself, the above rules do not
apply.  In the jargon, the two \rterm{plus} constraints within \rterm{R} remain
as \defn{delayed constraints}, which cannot do any work until more of their
variable arguments are replaced by constants (e.g., due to equality propagation
from a query, as above).

We can do better in this case with a library of additional rewrite rules that
implement standard axioms of arithmetic \cite{chr}, in particular the
associative law.  With these, \rterm{R} reduces to \rterm{plus(I,7,K)}, which is
a simpler description of this infinite relation.  Such rewrite rules are known
as idempotent \defn{constraint propagators}.  Other useful examples concerning
\rterm{plus} include \rRule{plus(0,J,K)}{K=J} and \rRule{plus(I,J,J)}{(I=0)},
since unlike the rules at the start of this section, they can make progress even
on a single \rterm{plus} constraint whose arguments include more than one
variable. Similarly, some
useful constraint propagators for the \rterm{lessthan} relation include
 \rRule{lessthan(J,J)}{0}; the transitivity rule
\rterm{lessthan(I,J)*lessthan(J,K)}$\rto\label{rr:lessthan_combine}$ \linebreak
\rterm{lessthan(I,J)*lessthan(J,K)*lessthan(I,K)}; and
\rterm{lessthan(0,I)*plus(I,J,K)}$\rto$%
\linebreak \rterm{lessthan(0,I)*plus(I,J,K)*lessthan(J,K)}.  The integer domain
can be \defn{split} by rules such as
\rRule{int(I)}{int(I)*(lessthan(0,I)+lessthan(I,1))} in order to allow case
analysis of, for example, \rterm{int(I)*myconstraint(I)}.  All of these rules
apply even if their arguments are variables, so they can apply early in a
reduction before other rewrites have determined the values of those variables.
Indeed, they can sometimes short-circuit the work of determining those
values.\looseness=-1

Like all rewrites, built-in rewrites \rRule{R}{S} must not change the denotation
of \rterm{R}: they ensure $\rSemE{R}{=}\rSemE{S}$ for all $\env$.  For example,
\rterm{lessthan(X,Y)*lessthan(Y,X)}$\rto^*$ \rterm{0} is semantics-preserving
because both forms denote the empty bag relation.

\subsection{Projection}\label{sec:proj}

Projection is implemented using the following rewrite rules.  The first two
rules make it possible to push the \rterm{proj(X,...)} operator down through the
sums and products of \rterm{R}, so that it applies to smaller subexpressions
that mention \rterm{X}:
\begin{termex}{}
proj(X,R+S) $\rtoleftright$ proj(X,R) + proj(X,S)  @\comm{distributivity over +}@
proj(X,R*S) $\rtoleftright\label{rr:push-in-proj}$ R*proj(X,S)  if X $\notin \vars(\rR)$ @\comm{see also the \texttt{R*}$\infty$ rule below}
\end{termex}
Using the following rewrite rules, we can then eliminate the projection operator
from smaller expressions whose projection is easy to compute.  (In other cases,
it must remain as a delayed operator.)  How are these rules justified?  Observe
that \rterm{proj(X,R)} in an environment $\env$ denotes the number of \rterm{X}
values that are consistent with $\env$'s binding of \rterm{R}'s other free
variables.  Thus, we may safely rewrite it as another expression that always
achieves the same denotation.
\begin{termex}{}
proj(X,(X=T)) $\rto$ $1$ if T $\in \theTerms$ and X $\notin \vars(\vT)$  @\comm{occurs check}@
proj(A,(A=sum(X,R))) $\rto$ $1$ if A $\notin \vars(\rR)$  @\comm{cardinality of an aggregated variable}@
proj(X,R) $\rto$ R*$\infty$  if X $\notin \vars(\rR)$  @\comm{cardinality of an unconstrained variable}@
proj(X,bool(X)) $\rto$ 2  @\comm{cardinality of a variable given a certain unary constraint}@
proj(X,int(X)) $\rto$ $\infty$    @\comm{cardinality of a variable given a certain unary constraint}@
proj(X,proj(Y,nand(X,Y))) $\rto$ 3 @\comm{card.\@ of a pair given a certain binary constraint}@
\end{termex}

\noindent As a simple example, let us project column \rterm{K} out of the table
\rterm{g(J,K)} from \cref{eq:ex:g-def}.

\hspace{-2ex}%
\begin{minipage}{.2\textwidth}
\begin{termex}{}
proj(K, g(J,K))
\end{termex}
\end{minipage}%
$\to$%
\begin{minipage}{.39\textwidth}
\begin{termex}{}
proj(K,((J=1) * (K=1)
      + (J=2) * (K=6)
      + (J=2) * (K=7)
      + (J=2) * (K=7)
      + (J=5) * (K=7)  ))
\end{termex}
\end{minipage}%
\hspace{-8ex}
$\to$
\begin{minipage}{.39\textwidth}
\begin{termex}{}
   ( (J=1) * proj(K, (K=1))
   + (J=2) * proj(K, (K=6))
   + (J=2) * proj(K, (K=7))
   + (J=2) * proj(K, (K=7))
   + (J=5) * proj(K, (K=7)) )
\end{termex}
\end{minipage}
\vspace{-.5\baselineskip}\qquad
\begin{termex}{}
                 $\to^*$ (J=1) + (J=2)*3 + (J=5)
\end{termex}

When multiple projection operators are used, we may push them down independently
of each other, since they commute:
\begin{termex}{}
proj(X, proj(Y, R)) $\to$ proj(Y, proj(X, R))
\end{termex}

\subsection{Aggregation} \label{sec:aggregation_rewrites}

The simple \rterm{count} aggregator from \cref{sec:syntax_semantics} is
implemented with the following rewrite rules, which resemble those for
\rterm{proj}:
\begin{termex}{}
M=count(R+S) $\rto$ proj(L,(L=count(R)) * proj(N,(N=count(S)) * plus(L,N,M)))
M=count(R*S) $\rto$ proj(L,(L=count(R)) * proj(N,(N=count(S)) * times(L,N,M))) if @$\vars(\rR) \cap \vars(\rS)=\emptyset$@
M=count(N) $\rto$ (M=N) if N $\in \theMults$
\end{termex}
In the first two rules, \rterm{L} and \rterm{N} are new bound variables
introduced by the right-hand side of the rule.  (When the rewrite rule is
applied, they will---as is standard---potentially be renamed to avoid capturing
free variables in the arguments to the left-hand side.)  They serve as temporary
registers.  The third rule covers the base case where the expression has been
reduced to a constant multiplicity: e.g.,
\begin{termex}{}
M=count(5=5) $\to$ M=count(1) $\to$ M=1
M=count(plus(I,J,J)*(I=5)) $\to$ M=count((I=0)*(I=5)) $\to^*$ M=count(0) $\to$ M=0
\end{termex}

The following rewrite rules implement \texttt{sum}. (The rules for other
aggregation operators are isomorphic.)  The usual strategy is to rewrite
\rterm{A=sum(X,R)} as a chain of \rterm{plus} constraints that maintain a
running total.  The following rules handle cases where \rterm{R} is expressed as
a union of 0, 1, or 2 bag relations, respectively.  (A larger union can be
handled as a union of 2 relations, e.g., \rterm{(Q+R)+S}.)
\begin{termex}{}
A=sum(X, 0) $\rto\label{rr:agg_sum1}$ (A=@\rid{sum}@)
A=sum(X, (X=T)) $\rto\label{rr:agg_sum2}$ (A=T) if T $\in \theTerms$ and X $\notin \vars(\vT)$   @\comm{occurs check}@
A=sum(X, R+S) $\rto\label{rr:agg_sum3}$ proj(B, (B=sum(X,R)) * proj(C, (C=sum(X,S)) * plus(B,C,A)))
\end{termex}

The second rule above handles only one of the base cases of 1 bag relation.  We
must add rules to cover other base cases, such as these:\footnote{As in
  \cref{sec:proj}, we could also include special rewrites for certain
  aggregations that have a known closed-form result, such as certain series
  sums.}
\begin{termex}{}
A=sum(X, (X=sum(Y,R))) $\rto$ (A=sum(Y,R)) if X $\notin \vars(\rR)$
A=sum(X, (X=min(Y,R))) $\rto$ (A=min(Y,R)) if X $\notin \vars(\rR)$
\end{termex}
Most important of all is this case, which is analogous to the second rule of
\cref{sec:proj} and is needed to aggregate over sum-of-products constructions:
\begin{termex}{}
A=sum(X,R*S) $\rto$ proj(B, sum_copies(R,B,A) * (B=sum(X,S)))  if X $\notin \vars(\rR)$
\end{termex}
Here, \rterm{sum_copies(M,B,A)} for \rterm{M} $\in\theMults$ constrains
\rterm{A} to be the aggregation of \rterm{M} $\in \theMults$ copies of the
aggregated value \rterm{B}.  The challenge is that in the general case we
actually have \rterm{sum_copies(R,B,A)}, so the multiplicity \rterm{M} may vary
with the free variables of \rterm{R}.  The desired denotational semantics are
\begin{itemize}
\item[] $\rSemE{ \rterm{sum_copies(R,B,A)} } = \textbf{ if } \rSemE{\rterm{R}} \cdot \rSemE{\rterm{B}} = \rSemE{\rterm{A}} \textbf{ then } 1 \textbf{ else } 0$
\item[] $\rSemE{ \rterm{min_copies(R,B,A)} } = \textbf{ if }
     (\rSemE{\rterm{R}}=0 \textbf{ and } \rSemE{\rterm{A}}=\rid{min})$ \\
     \hspace{5ex}$\textbf{ or } (\rSemE{\rterm{R}}>0 \textbf{ and } \rSemE{\rterm{A}}=\rSemE{\rterm{B}}) \textbf{ then } 1 \textbf{ else } 0$
\medskip
\item[] where $\rterm{R}\in\theRexprs$ and $\rterm{B},\rterm{A}\in\theTerms$
\end{itemize}
where we also show the interesting case of \rterm{min_copies(M,B,A)}, which is
needed to help define the \rterm{min} aggregator.  We can implement these by the
rewrite rules
\begin{termex}{}
$\!$sum_copies(R,B,A) $\!\rto\!$ proj(M,(M=count(R))*times(M,B,A)) @\comm{assumes \upshape{\rid{sum}$=0$}}@
$\!$min_copies(R,B,A) $\!\rto\!$ proj(M,(M=count(R))*((M=0)*(A=@\rid{min}@)+lessthan(0,M)*(A=B)))
\end{termex}

Identities concerning aggregation yield additional rewrite rules.  For example,
since multiplication distributes over $\sum$, summations can be merged and
factored via \rterm{(B=sum(I,R))*(C=sum(J,S))*}\rterm{times(B,C,A)} $\leftrightarrow$
\rterm{A=sum(K,R*S*times(I,J,K))} provided that
$\rterm{I}\in\vars(\rR),\rterm{J}\in\vars(\rS),\rterm{K}\notin\vars(\rterm{R*S})$,
and $\vars(\rR) \cap \vars(\rS)=\emptyset$.  Other distributive properties yield
more rules of this sort.  Moreover, projection and aggregation operators commute
if they are over different free variables.

     {%
     To conclude this section, we now attempt aggregation over infinite streams.  We
     wish to evaluate \rterm{A=exists(B,proj(I,peano(I)*myconstraint(I)*(B=true)))}
     to determine whether there exists any Peano numeral that satisfies a given
     constraint.  Here \rterm{exists} is the aggregation operator based on the binary
     \rterm{or} operation.

     \rterm{peano(I)} represents the infinite bag of Peano numerals, once we define a
     user constraint via the rewrite rule \rRule{peano(I)}{(I=zero) +
       proj(J,(I=s(J))*peano(J))}.  Rewriting \rterm{peano(I)} provides an
     opportunity to apply the rule again (to \rterm{peano(J)}).  After $k \geq 0$
     rewrites we obtain a representation of the original bag that explicitly lists
     the first $k$ Peano numerals as well as a continuation that represents all Peano
     numerals $\geq k$:
     \vspace{-\baselineskip}
     \newcommand{\tinycdots}{\!\cdot\!\cdot\!\cdot\!}
     \newcommand{\PeanoTemplate}[2]{\ensuremath{{\underbrace{\texttt{s(}\tinycdots\texttt{s(}}_{#1\text{ times}}}{\! #2}\texttt{)}\tinycdots\texttt{)}}}
     \begin{termex}{}
     $\to$ (I=zero) + $\cdots$ + (I=$\PeanoTemplate{k-1}{\texttt{zero}}$) + $\overbrace{\texttt{proj(J,(\vI{=}\!\PeanoTemplate{k-1}{\texttt{J}}))*peano(J)}}^{\text{continuation}}$
     \end{termex}%
     Rewriting the \rterm{exists} query over this $(k+1)$-way union results in a
     chain of $k$ \rterm{or} constraints.  If one of the first $k$ Peano numerals in
     fact satisfies \rterm{myconstraint}, then we can ``short-circuit'' the infinite
     regress and determine our answer without further expanding the continuation,
     thanks to the useful rewrite \rRule{or(true,C,A)}{(A=true)}, which can apply
     even while \rterm{C} remains unknown.

     In general, one can efficiently aggregate a function of the Peano numerals by
     alternating between expanding \rterm{peano} to draw the next numeral from the
     iterator, and rewriting the aggregating \rexpr to aggregate the value of the
     function at that numeral into a running ``total.''  If the running total ever
     reaches an absorbing element $a$ of the aggregator's binary operation---such as
     \rterm{true} for the \rterm{or} operation---then one will be able to simplify
     the expression to \rterm{A=}$a$ and terminate.  We leave the details as an
     exercise.

     }

\section{Translation of Dyna programs to \protect\rexprs} \label{sec:trans}

The translation of a Dyna program to a single recursive \rexpr can be performed
mechanically.  We will illustrate the basic construction on the small contrived
example below.  We will focus on the first three rules, which define \rterm{f}
in terms of \rterm{g}.  The final rule, which defines \rterm{g}, will allow us
to take note of a few subtleties.

\begin{dynaex}{}
f(X) += X*X.  @\label{line:translate_start}@
f(4) += 3.  @\label{line:f_rule_4}@
f(X) += g(X,Y).  @\label{line:translate_g}@
g(4*C,Y) += C-1 for Y > 99. @\label{line:translate_end}@
\end{dynaex}

Recall that a \dyna program represents a set of key-value pairs.
\rterm{is(Key,Val)} is the conventional name for the key-value relation.  The above program translates into the following user-defined constraint, which
recursively invokes itself when the value of one key is defined in terms of the
values of other keys.\looseness=-1

\begin{termex}{}
is(Key,Val) $\to$ (Val=sum(Result, @\comm{\texttt{sum} represents the \texttt{+=} aggregator}@
    proj(X, (Key=f(X))*times(X,X,Result) )  @\comm{\texttt{f(X) += X*X.}}@
   +        (Key=f(4))*(Result=3)        @\comm{\texttt{f(4) += 3.}}@
   +proj(X, (Key=f(X))*proj(Y,is(g(X,Y),Result))) )  @\comm{\texttt{f(X) += g(X,Y).}}@
   +proj(C, proj(Y, proj(Temp,(Key=g(Temp,Y))*times(4,C,Temp)) @\comm{\texttt{g(4*C,Y) +=\ }}@
                    *minus(C,1,Result)*lessthan(99,Y)   @\comm{\ldots \texttt{  C-1 for Y > 99.}}@
 ) * notnull(Val)  @\comm{\texttt{notnull} discards any pair whose \texttt{Val} aggregated nothing}@
\end{termex}

Each of the 4 Dyna rules translates into an \rexpr (as indicated by the comments
above) that describes a bag of (\rterm{Key},\rterm{Result}) pairs of ground
terms.  In each pair, \rterm{Result} represents a possible ground value of the
rule's body (the Dyna expression to the right of the aggregator \rterm{+=}), and
\rterm{Key} represents the corresponding grounding of the rule head (the term to
the left of the aggregator), which will receive \rterm{Result} as an aggregand.
Note that the same (\rterm{Key},\rterm{Result}) pair may appear multiple times.
Within each rule's \rexpr, we project out the variables such as \rterm{X} and
\rterm{Y} that appear locally within the rule, so that the \rexpr's free
variables are only \rterm{Key} and \rterm{Result}.\footnote{It is always legal
  to project out the local variables at the top level of the rule, e.g.,
  \rterm{proj(X,proj(Y,...))} for rule 3.  However, we have already seen rewrite
  rules that can narrow the scope of \rterm{proj(Y,...)} to a sub-\rexpr that
  contains all the copies of \rterm{Y}.  Here we display the version after these
  rewrites have been done.}

Dyna mandates in-place evaluation of Dyna expressions that have values
\cite{eisner-filardo-2011}.  For each such expression, we create a new local
variable to bind to the result.  Above, the expressions such as \rterm{X*X},
\rterm{g(X,Y)}, \rterm{4*C}, and \rterm{C-1} were evaluated using \rterm{times},
\rterm{is}, \rterm{times}, and \rterm{minus} constraints, respectively, and
their results were assigned to new variables \rterm{Result}, \rterm{Result},
\rterm{Temp}, and \rterm{Result}.  Importantly, \rterm{g(X,Y)} refers to a key
of the Dyna program itself, so the Dyna program translated into an \rterm{is}
constraint whose definition recursively invokes \rterm{is(g(X,Y),Result)}.

The next step is to take the bag union (\rterm{+}) of these 4 bag relations.
This yields the bag of all (\rterm{Key},\rterm{Result}) pairs in the program.
Finally, we wrap this combined bag in \rterm{Val=sum(Result,...)} to aggregate
the {\rterm{Result}}s for each \rterm{Key} into the \rterm{Val} for that key.
This yields a set relation with exactly one value for each key.  For
\rterm{Key=f(4)}, for example, the first and second rules each contribute
one \rterm{Result}, while the third rule contributes as many
{\rterm{Result}}s as the map has keys of the form
\rterm{g(4,Y)}.\footnote{\label{fn:infzeros}The reader should be able to see
  that the third Dyna rule will contribute infinitely many copies of \rterm{0},
  one for each \rterm{Y > 99}.  This is an example of multiplicity
  $\infty$. Fortunately, \rterm{sum_copies} (invoked when rewriting the
  \rterm{sum} aggregation operator) knows that summing any positive number of
  \rterm{0} terms---even infinitely many---will give \rterm{0}.}\looseness=-1

We use \rterm{sum} as our aggregation operator because all rules in this program
specify the \rterm{+=} aggregator.  One might expect \rterm{sum} to be based on
the binary operator that is implemented by the \rterm{plus} built-in, as
described before, with identity element \rid{sum}$\mbox{}=0$.  There is a
complication, however: in Dyna, a \rterm{Key} that has no {\rterm{Result}}s
should not in fact be paired with $\rterm{Val=0}$.  Rather, this \rterm{Key}
should not appear as a key of the final relation at all!  To achieve this, we
augment $\theFunctors$ with a new constant $\rNull$ (similar to Prolog's
\texttt{no}), which represents ``no results.''
We define \rid{sum}$\mbox{}=\rNull$, and we base \rterm{sum} on a modified
version of \rterm{plus} for which $\rNull$ is indeed the identity (so the
constraints \rterm{plus(null,X,X)} and \rterm{sum_copies(0,X,null)} are both
true for all \rterm{X}).  All aggregation operators in Dyna make use of $\rNull$
in this way.  Our \rterm{Val=sum(Result,...)} relation now obtains $\rNull$
(rather than \rterm{0}) as the unique \rterm{Val} for each \rterm{Key} that has
no aggregands. As a final step in expressing a Dyna program, we always remove
from the bag all $(\rterm{Key},\rterm{Val})$ pairs for which \rterm{Val=null},
by conjoining with a \rterm{notnull(Val)} constraint.  This is how we finally
obtain the \rexpr above for \rterm{is(Key,Val)}.

To query the Dyna program for the value of key \rterm{f(4)}, we can reduce the
expression \rterm{is(f(4),Val)}, using previously discussed rewrite rules.  We
can get as far as this before it must carry out its own query \rterm{g(4,Y)}:
\begin{termex}{}
proj(C, (C=sum(Result,proj(Y,is(g(4,Y),Result))))
        * plus(19,C,Val)                           ) * notnull(Val)
\end{termex}
where the local variable \rterm{C} captures the total contribution from rule 3,
and \rterm{19} is the total contribution of the other rules.  To reduce further,
we now recursively expand \rterm{is(g(4,Y),Result)} and ultimately reduce it to
\rterm{Result=0} (meaning that \rterm{g(4,Y)} turns out to have value \rterm{0}
for all ground terms \rterm{Y}).  \rterm{proj(Y,Result=0)} reduces to
{\rterm{(Result=0)*}}$\infty$---a bag relation with an infinite
multiplicity---but then {\rterm{C=sum(Result,(Result=0)*}}$\infty$\rterm{)}
reduces to \rterm{C=0} (via \rterm{sum_copies}, as \cref{fn:infzeros} noted).
The full expression now easily reduces to \rterm{Val=19}, the correct answer.

What if the Dyna program has different rules with different aggregators?  Then
our translation takes the form
\begin{termex}{}
is(Key,Val) $\rto\label{rr:is_call}$ Val=only(Val1,  (Val1=sum(Result, RSum))
                              + (Val1=min(Result, RMin))
                              + (Val1=only(Result, REq))
                              + ...                      ) * notnull(Val)
\end{termex}
where \rterm{RSum} is the bag union of the translated \rterm{+=} rules as in the
previous example, \rterm{RMin} is the bag union of the translated \rterm{min=}
rules, \rterm{REq} is the bag union of the translated \rterm{=} rules, and so
on.  The new aggregation operator \rterm{only} is based on a binary operator
that has identity \rid{only}$\mbox{}=\rNull$ and combines any pair of
non-$\rNull$ values into \rterm{error}.  For each \rterm{Key}, therefore,
\rterm{Val} is bound to the aggregated result \rterm{Val1} of the \emph{unique}
aggregator whose rules contribute results to that key.  If multiple aggregators
contribute to the same key, the value is \rterm{error}.\footnote{A Dyna program
  is also supposed to give an \rterm{error} value to the key \rterm{a} if the
  program contains both the rules \rterm{a = 3} and \rterm{a = 4}, or the rule
  \rterm{a = f(X)} when both \rterm{f(0)} and \rterm{f(1)} have values.  So we
  also used \rterm{only} above as the aggregation operator for \rterm{=} rules.}

A Dyna program may consist of multiple \emph{dynabases}
\cite{eisner-filardo-2011}. Each dynabase defines its own key-value mapping,
using aggregation over only the rules in that dynabase, which may refer to the
values of keys in other dynabases.  In this case, instead of defining a single
constraint \rterm{is} as an \rexpr, we define a different named constraint for
each dynabase as a different \rexpr, where each of the \rexprs may call any of
these named constraints.

\section{Related Work} \label{sec:related_work}

{\todocolor{purple}

\dyna is hardly the first programming language---or even the first logic programming language---to model its semantics in terms of relational expressions or to use term rewriting for execution.

Relational algebras provide the foundation of relational
databases \cite{codd1970relational}.  We require an extension from set relations to bag relations \citep[cf.][]{Green2009bag-semantics} in order to support aggregation, as aggregation considers the multiplicity of bag elements. Our particular formulation in \cref{sec:syntax_semantics} uses variable names rather than column indices; this allows us to construct and rearrange \rexprs without the technical annoyance of having to track and modify column indices.

Datalog has long been understood as an alternative mechanism for defining the sort of finite relations that are supported by relational databases \cite{datalog,datalog2}.  Indeed, it is possible to translate Datalog programs into a relational algebra that includes a fixpoint operator.  \citet{Bellia90} and \citet{Arias17} generalize some of these translation constructions to pure logic programming, though without aggregation or built-ins. To attain the full power of relational algebras, Datalog is often extended to support aggregation and negation (using ``stratified'' programs to control the interaction of these operations with recursion). Implementations of Datalog often use a forward-chaining (or ``semi-naive bottom-up'') strategy that materializes all relations in the user-defined program \cite{ullman-1988}, and this was used in the first version of \dyna \cite{eisner-goldlust-smith-2005}.  

Materialization is a sort of brute-force strategy that
is guaranteed to terminate (for stratified programs) when all relations are finite, but which is not practical for very large or infinite relations.  Thus, our goal in this paper was to make it possible to manipulate intensional descriptions of possibly infinite relations.  The work that is closest to our approach is constraint logic programming \cite{Jaffar:CLPsurvey}, which extends Prolog's backtracking execution (SLD resolution) so that the environment in which a subgoal is evaluated includes not only variable bindings from Prolog unification, but also a ``store'' of \defn{delayed constraints} (so called because they have not yet been rewritten into simple unification bindings as would be required by SLD resolution).  This conjunction of unification binding constraints and delayed constraints on a common set of variables is continually rewritten using constraint handling rules \cite{fruehwirth-1998} such as simplification and propagation.  Our pseudocode in \cref{sec:pseudocode} shares this notion of an environment with constraint logic programming, and applies our rewrite rules from \cref{sec:rewrite_rules} in essentially the same way.  One difference is that our pseudocode is breadth-first---it rewrites disjuncts in parallel rather than backtracking.  This increases the memory footprint, but sometimes makes it possible to short-circuit an infinite regress that is provably irrelevant to the final answer.  (Our rewrite rules could be applied in any order, of course.)  Another difference is that our \rexpr formalism supports the construction of new relations from old ones by aggregation and projection, as required by the Dyna language.  To our knowledge, these extensions to expressivity have not been proposed within constraint logic programming (although \citet{ross-et-al-1998} do consider aggregation constraints on first-class bag-valued variables).

Of course, functional programming is also based on term rewriting (e.g., reduction strategies for the $\lambda$-calculus), and term rewriting has figured in attempts to combine functional and logic programming. The Curry language~\cite{Hanus:1997:UCM:263699.263710,Hanus07ICLP} rewrites terms that denote computable functions, but supports the Prolog-like ability to call a function on variables that have not yet been bound.  This motivates a rewriting technique called narrowing~\cite{ANTOY2010501} that can result in nondeterministic computation (e.g., backtracking) and is related to unfolding (inlining) in logic programming.  Whereas Curry simplifies terms that denote arbitrary functions (potentially higher-order functions), our proposed system for \dyna rewrites terms that specifically denote bag relations (i.e., functions from ground tuples to multiplicities).  We remark that Curry appears to have committed to a specific rewrite ordering strategy resembling Haskell's fixed execution order.  While we give a specific strategy for rewriting \dyna in \cref{sec:pseudocode}, we have also explored the possibility of \emph{learning} execution strategies for \dyna that tend to terminate quickly on a given workload \cite{mapl2017}.

\section{Conclusion}

{%

We have shown how to algebraically represent the bag-relational semantics of any program written in a declarative logic-based language like Dyna.  A query against a program can be evaluated by joining the query to the program and simplifying the resulting algebraic expression with appropriate term rewriting rules.  It is congenial that this approach allows evaluation to flexibly make progress, propagate information through the expression, exploit arithmetic identities, and remove irrelevant subexpressions rather than wasting possibly infinite work on them.

In ongoing work, we are considering methods for practical interpretation and compilation of rewrite systems.  Our goal is to construct an evaluator that will both perform static analysis and ``learn to run fast'' \cite{mapl2017} by constructing or discovering reusable strategies for reducing expressions of frequently encountered sorts (i.e., polyvariant specialization, including memoization, together with programmable rewrite strategies).
}

\bibliographystyle{acmtrans}
\bibliography{bibfile}

\begin{thebibliography}{}

\bibitem[\protect\citeauthoryear{Antoy}{Antoy}{2010}]{ANTOY2010501}
{\sc Antoy, S.} 2010.
\newblock Programming with narrowing: A tutorial.
\newblock {\em Journal of Symbolic Computation\/}~{\em 45,\/}~5, 501 -- 522.
\newblock Symbolic Computation in Software Science.

\bibitem[\protect\citeauthoryear{Arias, Lipton, and Mariño}{Arias
  et~al\mbox{.}}{2017}]{Arias17}
{\sc Arias, E. J.~G.}, {\sc Lipton, J.}, {\sc and} {\sc Mariño, J.} 2017.
\newblock Constraint logic programming with a relational machine.

\bibitem[\protect\citeauthoryear{Bellia and Occhiuto}{Bellia and
  Occhiuto}{1990}]{Bellia90}
{\sc Bellia, M.} {\sc and} {\sc Occhiuto, M.~E.} 1990.
\newblock C-expressions: a variable-free calculus for equational logic
  programming.

\bibitem[\protect\citeauthoryear{Ceri, Gottlob, and Tanca}{Ceri
  et~al\mbox{.}}{1989}]{datalog}
{\sc Ceri, S.}, {\sc Gottlob, G.}, {\sc and} {\sc Tanca, L.} 1989.
\newblock What you always wanted to know about {Datalog} (and never dared to
  ask).
\newblock In {\em IEEE Transactions on Knowledge and Data Engineering}.

\bibitem[\protect\citeauthoryear{Clocksin and Mellish}{Clocksin and
  Mellish}{1984}]{prolog2}
{\sc Clocksin, W.~F.} {\sc and} {\sc Mellish, C.~S.} 1984.
\newblock {\em Programming in Prolog}.
\newblock Springer-Verlag.

\bibitem[\protect\citeauthoryear{Codd}{Codd}{1970}]{codd1970relational}
{\sc Codd, E.~F.} 1970.
\newblock A relational model of data for large shared data banks.
\newblock {\em Communications of the ACM\/}~{\em 13,\/}~6.

\bibitem[\protect\citeauthoryear{Colmerauer and Roussel}{Colmerauer and
  Roussel}{1996}]{prolog}
{\sc Colmerauer, A.} {\sc and} {\sc Roussel, P.} 1996.
\newblock {\em The Birth of Prolog}.
\newblock Association for Computing Machinery, Chapter~7.

\bibitem[\protect\citeauthoryear{Dijkstra}{Dijkstra}{1959}]{dijkstra-1959}
{\sc Dijkstra, E.~W.} 1959.
\newblock A note on two problems in connexion with graphs.
\newblock {\em Numerische Mathematik\/}~{\em 1}, 269--271.

\bibitem[\protect\citeauthoryear{Eisner and Filardo}{Eisner and
  Filardo}{2011}]{eisner-filardo-2011}
{\sc Eisner, J.} {\sc and} {\sc Filardo, N.~W.} 2011.
\newblock Dyna: Extending {D}atalog for modern {AI}.
\newblock In {\em Datalog Reloaded}, {O.~de~Moor}, {G.~Gottlob}, {T.~Furche},
  {and} {A.~Sellers}, Eds. Lecture Notes in Computer Science. Springer.

\bibitem[\protect\citeauthoryear{Eisner, Goldlust, and Smith}{Eisner
  et~al\mbox{.}}{2005}]{eisner-goldlust-smith-2005}
{\sc Eisner, J.}, {\sc Goldlust, E.}, {\sc and} {\sc Smith, N.~A.} 2005.
\newblock Compiling comp ling: Weighted dynamic programming and the {D}yna
  language.
\newblock In {\em Proceedings of Human Language Technology Conference and
  Conference on Empirical Methods in Natural Language Processing (HLT-EMNLP)}.
  Vancouver, 281--290.

\bibitem[\protect\citeauthoryear{Fr\"uhwirth}{Fr\"uhwirth}{1998}]{fruehwirth-1998}
{\sc Fr\"uhwirth, T.} 1998.
\newblock Theory and practice of constraint handling rules.
\newblock {\em The Journal of Logic Programming\/}~{\em 37,\/}~1.

\bibitem[\protect\citeauthoryear{Frühwirth}{Frühwirth}{1998}]{chr}
{\sc Frühwirth, T.} 1998.
\newblock Theory and practice of constraint handling rules.
\newblock {\em The Journal of Logic Programming\/}~{\em 37,\/}~1.

\bibitem[\protect\citeauthoryear{Gallaire, Minker, and Nicolas}{Gallaire
  et~al\mbox{.}}{1984}]{datalog2}
{\sc Gallaire, H.}, {\sc Minker, J.}, {\sc and} {\sc Nicolas, J.-M.} 1984.
\newblock Logic and databases: A deductive approach.
\newblock {\em ACM Comput. Surv.\/}~{\em 16,\/}~2.

\bibitem[\protect\citeauthoryear{Green}{Green}{2009}]{Green2009bag-semantics}
{\sc Green, T.~J.} 2009.
\newblock Bag semantics.
\newblock In {\em Encyclopedia of Database Systems}, {L.~LIU} {and} {M.~T.
  {\"O}ZSU}, Eds. Springer.

\bibitem[\protect\citeauthoryear{Hanus}{Hanus}{1997}]{Hanus:1997:UCM:263699.263710}
{\sc Hanus, M.} 1997.
\newblock A unified computation model for functional and logic programming.
\newblock In {\em Proceedings of the 24th ACM SIGPLAN-SIGACT Symposium on
  Principles of Programming Languages}. POPL '97. ACM, New York, NY, USA.

\bibitem[\protect\citeauthoryear{Hanus}{Hanus}{2007}]{Hanus07ICLP}
{\sc Hanus, M.} 2007.
\newblock Multi-paradigm declarative languages.
\newblock In {\em Proceedings of the International Conference on Logic
  Programming (ICLP 2007)}. Springer LNCS 4670, 45--75.

\bibitem[\protect\citeauthoryear{Jaffar and Maher}{Jaffar and
  Maher}{1994}]{Jaffar:CLPsurvey}
{\sc Jaffar, J.} {\sc and} {\sc Maher, M.~J.} 1994.
\newblock Constraint {L}ogic {P}rogramming: A survey.
\newblock {\em J. Logic Program.\/}~{\em 19,20}, 503--581.

\bibitem[\protect\citeauthoryear{Ross, Srivastava, Stuckey, and Sudarshan}{Ross
  et~al\mbox{.}}{1998}]{ross-et-al-1998}
{\sc Ross, K.~A.}, {\sc Srivastava, D.}, {\sc Stuckey, P.~J.}, {\sc and} {\sc
  Sudarshan, S.} 1998.
\newblock Foundations of aggregation constraints.
\newblock {\em Theoretical Computer Science\/}~{\em 193,\/}~1, 149--179.

\bibitem[\protect\citeauthoryear{Ullman}{Ullman}{1988}]{ullman-1988}
{\sc Ullman, J.~D.} 1988.
\newblock {\em Principles of Database and Knowledge-Base Systems}. Vol.~1.
\newblock Computer Science Press.

\bibitem[\protect\citeauthoryear{Vieira, Francis-Landau, Filardo, Khorasani,
  and Eisner}{Vieira et~al\mbox{.}}{2017}]{mapl2017}
{\sc Vieira, T.}, {\sc Francis-Landau, M.}, {\sc Filardo, N.~W.}, {\sc
  Khorasani, F.}, {\sc and} {\sc Eisner, J.} 2017.
\newblock Dyna: Toward a self-optimizing declarative language for machine
  learning applications.
\newblock In {\em Proc. of the ACM SIGPLAN Workshop on Machine Learning and
  Programming Languages}.

\end{thebibliography}

\newpage
\appendix

\section{How \protect\rexprs Manage Infinite Relations} \label{sec:neural_ex}

%

%

%
%
%
%

%
%
%
%
%
%
%

%
%
%
%
%
%
%
%
%
%
%

%
%
%
%

%
%

%
%
%
%
%
%
%
%
%

%
%
%
%
%
%
%
%
%
%
%
%
%
%
%
%
%
%
%

%
%
%
%
%
%
%
%
%
%

%

%

%
%
%
%
%
%
%
%

%
%
%
%

%
%
%
%
%
%
%
%
%
%
%

%
%
%
%
%

%

%

Here we illustrate how to successfully simplify \rexprs that involve infinite relations, as in our two examples of \cref{sec:dynaex}: all-pairs shortest path and convolutional neural networks.  We focus on how our term rewriting approach is able to handle the neural network example (lines~\ref{line:conv_start}--\ref{line:output_weight}).  The shortest-path example succeeds for similar reasons, and we leave it as an exercise for the reader.\footnote{The answer appears on our talk slides from WRLA 2020 (see \cref{footnote:pubinfo}).}

Recall that the neural network defined in \cref{sec:dynaex} defines an \emph{infinite} number of \rterm{edge}\!s between input units and hidden units, not all of which are used in the case of a finite image.\footnote{Our approach can also attempt to answer queries even when the image is an infinite image that is specified by rule, such as an all-white image or a tessellation of the plane. Depending on the specification of the image, the resulting \rexpr may or may not simplify to yield a simple numerical answer for (say) the value of \dynainline{out(output(kitten))}.  We do not consider such cases in this appendix.}  To be able to successfully answer queries against this program, we must avoid materializing the infinite \rterm{edge} relation.

First, we translate the program at lines~\ref{line:conv_start}--\ref{line:output_weight}
 \dyna into a user-defined constraint, \mbox{$\rterm{is(Key,Val)} \to \cdots$}.  For readability, we show only the specialization of this constraint to cases where \rterm{Key} is an \rterm{edge}:
 \begin{termex}{}recur
is(edge(Arg1,Arg2),Result) $\to$ proj(X, proj(Y, proj(DX, proj(DY, proj(XX, proj(YY,
    (Arg1=input(X,Y))*(Arg2=hidden(XX, YY))* @\comm{the nested \texttt{Key} pattern has been decomposed}@
    @\comm{into equality constraints}@
    plus(X,DX,XX)*plus(Y,DY,YY)*  @\comm{arithmetic \texttt{+} in \dyna becomes \texttt{plus} in the \rexpr;}@
    @\comm{new variables \texttt{XX} and \texttt{YY} are introduced for the results}@
    is(weight_conv(DX, DY), Result)  @\comm{recurse to another branch of the \texttt{is} constraint (below)}@
)))))) + proj(XX, proj(YY, proj(Property, 
  (Arg1=hidden(XX, YY))*(Arg2=output(Property))*
  is(weight_output(Property), Result) @\comm{recurse to another branch of the \texttt{is} constraint (below)}@
)))
\end{termex}
The first summand describes the input-to-hidden edges, and the second summand describes the hidden-to-output edges.  The specializations of the user-defined constraint to the weight parameters would look like this (recall that the parameters were set randomly):
\begin{termex}{}
is(weight_conv(DX, DY), Result) $\to$ (  @\comm{\texttt{weight\_conv} is a table similar to \cref{sec:finite_materialized}}@
  (DX=-4)*(DY=-4)*(Result = .123) +
  (DX=-4)*(DY=-3)*(Result = .173) +
  (DX=-4)*(DY=-2)*(Result = .281) +
  $\vdots$
  (DX=4)*(DY=4)*(Result = .971) )
  
is(weight_output(Property), Result) $\to$ ( @\comm{\texttt{weight\_output} table}@
  (Property=kitten)*(Result=.777) +
  (Property=puppy)*(Result=.643) +
  $\vdots$
  (Property=zebra)*(Result=.912) )
\end{termex}

The right-hand side of the first rule can be further rewritten by replacing the call to
\rterm{weight\_conv} with its definition.  Then we can use the rewrite rules from \cref{sec:proj} that to lift the resulting disjunction out of the scope of the \rterm{proj} operator.
This gives us the following answer to the query \dynainline{is(edge(Arg1, Arg2), Result)}, which cannot be further simplified:
\newpage
\begin{termex}{}
is(edge(Arg1, Arg2), Result) $\to^*$ proj(X, proj(Y, proj(XX, proj(YY,
    (Arg1=input(X,Y))*(Arg2=hidden(XX,YY))*
    plus(X,-4,XX)*plus(Y,-4,YY)*
    (Result=.123)
    )))) + proj(X, proj(Y, proj(XX, proj(YY,
    (Arg1=input(X,Y))*(Arg2=hidden(XX,YY))*
    plus(X,-4,XX)*plus(Y,-3,YY)*
    (Result=.173)
    )))) + $\cdots$ @\comm{Additional branches of the disjunction omitted}@
\end{termex}

This answer still defines infinitely many edges, where the variables \rterm{X} and \rterm{XX} (similarly \rterm{Y} and \rterm{YY}) are related by \rterm{plus} constraints.  

If we were rewriting the above \rexpr in the context of a larger computation such as a query for \dynainline{out(output(kitten))} on a particular image, then we would be able to specialize it further against the finite set of input pixels.  In this
case, suppose that only a single input unit  \rterm{Arg1=input(0,0)} appears as a key in the \rterm{is} relation.  When
combined as \rterm{is(edge(Arg1, Arg2), Result)*(Arg1=input(0,0))}, this will eliminate the disjunctive branches of \texttt{weight\_output} and 
set the values of \rterm{X=0} and \rterm{Y=0} for all of the \texttt{weight\_conv} rules.  This will then allow for all of
the \rterm{plus(X,-4,XX)} and \rterm{plus(Y,-4,YY)} expressions above to run.
Further propagation of the values of these constraints will eventually determine
the value of \rterm{Arg2}, identifying the finite number of edges which are
actually required to run this expression:
\begin{termex}{}
is(edge(Arg1,Arg2),Result)*(Arg1=input(0,0)) $\to$ (
  (Arg2=output(-4,-4))*(Result=.123) +
  (Arg2=output(-4,-3))*(Result=.173) +
  $\cdots$ )
\end{termex}

\section{\protect\rexpr Solver Implementation Details}
\label{sec:exec_with_r}

\newcommand{\renv}{\ensuremath{\mathcal{C}}\xspace}

\newcommand{\SimplifyComplete}[0]{SimplifyComplete}
\newcommand{\SimplifyFast}[0]{SimplifyFast}
\newcommand{\SimplifyCompletePartition}[0]{SimplifyCompleteMain}
\newcommand{\findConj}[0]{GatherEnvironment}
\newcommand{\findDisj}[0]{GatherBranches}

%
%
%
%
%
%
%
%
%
%

%
%

%

%

%
%
%
%

%

%
%
%
%
%
%
%

%

%
%
%
%
%
%
%
%
%

%
%
%
%
%

%
%
%

%
%
%
%
%
%
%

%
%
%
%
%
%
%
%
%

%
%
%

%
%

%
%

%

%

%
%
%
%
%

%
%
%

%
%
%
%
%
%

%
%
%

%
%
%
%

%

%
%
%
%
%
%
%
%
%
%
%
%
%
%
%
%

%

%
%
%

%

%
%
%
%

%
%
%
%
%
%
%
%

%
%
%
%
%
%
%
%

%
%
%
%
%
%
%
%
%
%
%

%
%
%
%
%
%
%
%
%

%
%
%
%
%
%
%
%
%
%
%
%
%

%
%
%
%
%
%
%
%
%
%
%

%
%
%
%
%
%
%

%
%
%
%
%
%
%
%
%
%
%

%
%
%
%
%
%
%
%
%
%
%
%

%
%
%
%
%
%
%
%
%
%
%
%
%
%
%

%
%

%
%
%
%

%
%
%
%
%

%

%

%
%
%
%
%

%
%
%

%

%

%

%
%
%
%
%
%
%
%
%
%
%
%
%
%
%
%
%
%

%
%
%
%

%

%

%
%
%

%
%
%

%

%
%
%

%
%
%

%

%
%
%
%

%
%

%
%
%

%
%

%

%
%

%
%
%
%
%
%

%

%
%

%
%
%
%
%
%
%
%
%
%
%
%
%
%
%
%
%
%
%

%
%
%
%
%

\subsection{Simplification in an environment}

Our rewriting procedure is based around \emph{simplifying} a \rexpr until it
reaches a normal form (meaning that there are no more rewrites which can be
applied).

Scanning an entire \rexpr and identifying all available rewrites could be
expensive if we were to directly apply the rules presented in this paper.  One
reason is that many of the rules can only apply once two distant constraints
have been brought together into a conjunction \rterm{R*S}.  For example, a
reduction like
$\rterm{(X=2)*}\cdots\rterm{*(plus(X,3,Y)+(Z=9))}\to^*\rterm{(X=2)*}\cdots\rterm{*((Y=5)+(Z=9)}$
requires quite a lot of rearrangement to create a copy of \rterm{X=2} (using the
rewrite \rterm{R} $\to$ \rterm{R*R}), move this copy rightward
(using associative and commutative rewrites), and push it down into the
disjuncts (using a distributive rewrite), finally obtaining a subterm
\rterm{(X=2)*plus(X,3,Y)} that can be rewritten as \rterm{(X=2)*(Y=5)}.  Now we
have \rterm{(X=2)*}$\cdots$\rterm{*((X=2)*(Y=5)+(X=2)*(Z=9))}, and many of these
steps must be carried out in reverse to consolidate the extra copies of
\rterm{X=2} back into the original.

Our top-down approach will instead implicitly percolate constraints to where
they are needed by maintaining an \defn{environment}---a bag of currently
available constraints conjunctive, essentially the same as the constraint store
used in constraint logic programming.  In the example above, when the rewriting
engine encounters \rterm{plus(X,3,Y)}, it will be able to look up the value of
\rterm{X} in the current environment $\renv$.  This environment is passed by
reference through the operational semantics pseudocode (below) in much the same
way as the environment of \cref{sec:syntax_semantics} was threaded through the
denotational semantics definitions, although the two notions of environment are
distinct and have different types.

An important benefit of using an environment is that we can eliminate reversible
rules that were presented in the main paper (associativity, commutativity,
distributivity, constraint duplication), whose reversibility prevented the
existence of normal forms. In effect, we now have a composite rule
something like ``rewrite \rterm{plus(X,3,Y)} $\to$ \rterm{Y=5} provided that
\rterm{X=2} must also be true,'' which implicitly invokes whatever
associative/commutative/distributive rewrites are necessary to percolate a copy
of \rterm{X=2} to the position where it can license this composite rule.  We
need not explicitly invoke the reversible rules.

We define a \Call{Simplify}{} operator which takes a \rexpr \rterm{R} and
returns its normal form.  The operator is also given the current environment as
a second argument, as a place to find and store unordered conjunctive
constraints outside of \rterm{R} itself.

Internally, \Call{Simplify}{} makes use of two operators \Call{\SimplifyFast}{}
and \Call{\SimplifyComplete}{}.  \Call{\SimplifyFast}{} allows the system to
quickly identify rewrites that are useful, but which require a \emph{minimal}
amount of work to identify.  It works by using the environment to track ground
assignments to variables.  This allows a constraint like \rterm{plus(X,3,Y)} to
quickly check the environment to see whether \rterm{X} or \rterm{Y} is currently
bound to a number without having to scan or modify the entire \rexpr.
\Call{\SimplifyComplete}{} is similar, but it identifies \emph{all} possible
rewrites that can currently be applied (not just the fast ones).  Thus, it uses
an enhanced environment that can also track other types of constraints, similar
to the constraint store in ECLiPSe.  The enhanced environment may now include
structural equality constraints on non-ground terms, other built-in constraints
such as \rterm{plus} and \rterm{lessthan}, user-defined constraints, and even
constraints such as \rterm{proj(X,}$\cdots$\rterm{)} and
\rterm{A=sum(}{}$\cdots$\rterm{)}.    This allows the system to
rapidly identify the contradiction in an expression such as
\rterm{lessthan(A,B)*}$\cdots$\rterm{*lessthan(B,A)}$\to$\rterm{ 0} without
using explicit associative or commutative rewrites.
The enhanced environment does not include disjunctive constraints of the form \rterm{R+S}.

\subsection{Pseudocode of \textsc{Simplify}} \label{sec:pseudocode}

To simplify an \rexpr to a normal form, the system starts by applying
\Call{\SimplifyFast}{} repeatedly until it is unable to make any more changes.
The system will then try a single application of \Call{\SimplifyComplete}{} in case there are any additional rewrites that can be applied at this point.  Note: the entry point for \Call{\SimplifyComplete}{} is \Call{\SimplifyCompletePartition}{}.  If so, it returns to \Call{\SimplifyFast}{} and repeats the process.

\begin{algorithmic}[1]
  \Function{Simplify}{\protect\rterm{R}}%
    \State \renv $\gets \emptyset$ \Comment{Construct an empty environment to start}
    \State \rterm{R} $\gets$ \Call{UniqifyVars}{\protect\rterm{R}} \Comment{Ensure that all of the variables in the \rexpr have unique names}
    \Loop \Comment{Outer loop tries \textsc{\SimplifyComplete} only when there are no more fast rewrites available.}
      \Loop \Comment{Inner loop performs faster rewrites using \textsc{\SimplifyFast}}
        \State \rterm{R'} $\gets$ \Call{\SimplifyFast}{\protect\rterm{R}, \renv}
        \If{\rterm{R} == \rterm{R'}}
          \textbf{break} \Comment{\textsc{\SimplifyFast} has finished rewriting}
        \EndIf
        \State \rterm{R} $\gets$ \rterm{R'}
      \EndLoop
      \State \rterm{R'} $\gets$ \Call{\SimplifyCompletePartition}{\protect\rterm{R}, \renv} %
      \If{\rterm{R} == \rterm{R'}}
        \textbf{break} \Comment{\textsc{\SimplifyComplete} has finished rewriting}
      \EndIf
      \State \rterm{R} $\gets$ \rterm{R'}
    \EndLoop
    \State \Return \renv\ \rterm{* R}  \Comment{The \rexpr has reached a normal form, return to caller}
  \EndFunction
\end{algorithmic}

\Call{UniqifyVars}{\protect\rterm{R}} is a preprocessing step that
ensures that distinct variables within \rterm{R} have unique names.  In practice, it accomplishes this by giving them distinct integer subscripts.  In particular, in an operator sub-expression of the form \rterm{proj(X,S)} or \rterm{sum(X,S)}, an unused integer $i$ is chosen, and the first argument \rterm{X} as well as all copies of \rterm{X} that are bound by it are replaced with $\rterm{X}_i$.

The invariant of having unique names will be preserved by the rest of the \Call{Simplify}{} procedure.  In particular, when a rewrite rule introduces new variables (e.g., when expanding an aggregation or inlining a user-defined constraint), these variables are assigned unused names.  In practice, the variable names are augmented with unused integers as subscripts.

An environment \renv consists primarily of an array or hash map that maps key $i$ to a list of the constraints in \renv that constrain the variable subscripted with $i$.  Thanks to the unique-name invariant, distinct variables such as $\rterm{X}_5$ and $\rterm{X}_8$ can appear in the same environment.

\Call{\SimplifyFast}{} performs some simple rewrites in a single pass over the \rexpr, matching against nodes of the \rexpr syntax tree.  The pseudocode shown here is incomplete (it omits some simple rewrites that are in fact handled by our \Call{\SimplifyFast}{}, such as multiplicity arithmetic from \cref{sec:mult_arith}).  However, it gives the general flavor of how rewriting with an
environment works.  The environment is passed by reference and is destructively modified.
At the top level, \Call{\SimplifyFast}{} is called on \rterm{R} and an empty environment \renv.  It returns \rterm{R'} and destructively adds some constraints into \renv, which are no longer necessarily enforced by \rterm{R'}.  Thus, \rterm{R} is denotationally equivalent to \renv\rterm{*R'} (where \renv is interpreted as the product of the constraints in \renv), and $\rterm{R} \to^* \renv\rterm{*R'}$ under the rewrite rules of the main paper.
\begin{algorithmic}[1]
  \Function{\SimplifyFast}{\protect\rterm{R}, \renv}
    \If{\rterm{R} \textbf{ matches } \rterm{(X=x)}{} \textbf{ with } $\rterm{x}\in\theGround$} \Comment{Handle equality constraints (with ground values).  \cref{rr:equality_prop}}
      \If{\rterm{X} $\in$ \renv} \label{alg:line:equality_compare}  \Comment{Check if \vX's value in \renv matches the constraint {\tt x}}
        \If{{\renv}[\rterm{X}{}] == x}  \Comment{We use ``{\renv}[\rterm{X}{}]'' as a shorthand for ``\rterm{(X=x)} $\in$ \renv'' that returns ``\rterm{x}''}
          \State \Return \rterm{1} \Comment{Successful unification}
        \Else
          \State \Return \rterm{0}
        \EndIf
      \Else
        \State {\renv}[\rterm{X}{}] $\gets$ x \label{alg:absorb-into-env} \Comment{\renv is modified in place}
        \State \Return \rterm{1} \label{alg:line:equality_compare2}
      \EndIf
    \ElsIf{\rterm{R} \textbf{ matches } \rterm{(X=Y)} \textbf{ and } \rterm{Y} $\in$ \renv}
      \LineComment{Same as lines \ref{alg:line:equality_compare} through \ref{alg:line:equality_compare2}, but setting  \emph{\rterm{x}={\renv}[\rterm{Y}{}]}}
    \ElsIf{\rterm{R} \textbf{ matches } \rterm{(X=X)}}
      \State \Return \rterm{1}  \Comment{Vacuously true}
    \ElsIf{\rterm{R} \textbf{ matches } $\rterm{(X=f(U}_1\rterm{,}\ldots\rterm{,U}_n$\rterm{))}} \Comment{Structural unification between ground terms.  \cref{rr:struct_unify1,rr:struct_unify2}}
      \If{\rterm{X} $\in$ \renv} \Comment{In the case that X is a ground value}
        \State $x$ $\gets$ {\renv}[\rterm{X}{}]
        \State \Return $\rterm{(U}_1\rterm{=}x_1\rterm{)*}\cdots\rterm{*(U}_n\rterm{=}x_n\rterm{)}$ \Comment{Unify all of the arguments of the structure}
      \ElsIf{$\rterm{U}_1$ $\in$ \renv \textbf{ and } $\cdots$ \textbf{ and } $\rterm{U}_n$ $\in$ \renv} \Comment{All of $U_i$ are ground values}
        \State $u_1 \gets $ {\renv}[$\rterm{U}_1$] ; $\cdots$ ; $u_n \gets $ {\renv}[$\rterm{U}_n$]
        \State \Return $\rterm{(X=f(}u_1\rterm{,}\ldots\rterm{,}u_n\rterm{))}$ \Comment{Construct a term from the values of $u_i$}
      \Else
        \State \Return \rterm{R} \Comment{Neither side is ground, no change in the returned value}
      \EndIf
    \ElsIf{\rterm{R} \textbf{ matches } \rterm{Q*S}{}}  \Comment{Conjunctive \rexprs}
      \State \rterm{Q'} $\gets$ \Call{\SimplifyFast}{\protect\rterm{Q}, \renv}
      \If{\rterm{Q'} \textbf{ matches } \rterm{0}}
        \Return \rterm{0}  \Comment{The first conjunct has failed, so it can stop processing early. Implements \cref{rr:zero_mult}}
      \EndIf
      \State \rterm{S'} $\gets$ \Call{\SimplifyFast}{\protect\rterm{S}, \renv}
      \If{\rterm{S'} \textbf{ matches } \rterm{0}}
        \Return \rterm{0} \Comment{Implements \cref{rr:zero_mult}}
      \EndIf
      \If{\rterm{Q'} \textbf{ matches } \rterm{1}}
        \Return \rterm{S'} \Comment{Implements \cref{rr:one_mult}}
      \EndIf
      \If{\rterm{S'} \textbf{ matches } \rterm{1}}
        \Return \rterm{Q'} \Comment{Implements \cref{rr:one_mult}}
      \EndIf
      \If{\rterm{S'} $\in \theMults$ \textbf{ and } \rterm{Q'} $\in \theMults$} \Comment{Product of multiplicities as in \cref{rr:multiply_mult}}
        \State \rterm{L} $\gets$ \rterm{S'} * \rterm{Q'}
        \State \Return \rterm{L}
      \EndIf
      \State \Return \rterm{Q'*S'}
    \ElsIf{\rterm{R} \textbf{ matches } \rterm{Q+S}{}} \Comment{Disjunctive \rexprs}
      \LinesComment{Below, we copy the environment so that it can be modified separately by each recursive call. These could be lazy copies (copy on write), or managed via a call stack with undo operations as in Prolog.}
      \State $\renv_Q$ $\gets$ \Call{Copy}{\renv}
      \State $\renv_S$ $\gets$ \Call{Copy}{\renv}
      \State \rterm{Q'} $\gets$ \Call{\SimplifyFast}{\protect\rterm{Q}, $\renv_Q$}
      \State \rterm{S'} $\gets$ \Call{\SimplifyFast}{\protect\rterm{S}, $\renv_S$}
      \If{\rterm{Q'} \textbf{ matches } \rterm{0}}  \Comment{Remove \rterm{Q'} as it is now an empty branch}
        \State $\renv \gets \renv_S$ \Comment{Copy the environment of non-empty branch}
        \State \Return\rterm{S'}
      \EndIf
      \If{\rterm{S'} \textbf{ matches } \rterm{0}}
        \State $\renv \gets \renv_Q$
        \State \Return \rterm{Q'}
      \EndIf
      \LinesComment{Below, any constraints that are shared between $\renv_Q$ and $\renv_S$ are propagated to $\renv$.  Recall that \textsc{\SimplifyFast}'s environment only contains ground variable assignments, so this is only checking equalities between the same key for the variables.}
      \State \renv $\cup$= $\renv_Q$ $\cap$ $\renv_S$\label{alg:stash-start} \Comment{Implement \cref{rr:distributivity} for ground assignments only}
      \LinesComment{Anything in $\renv_Q$ and $\renv_S$ that could not be merged into \renv is added back to the \rexpr returned.}
      \If{\rterm{Q'} $\in\theMults$ \textbf{ and } \rterm{S'} $\in\theMults$ \textbf{ and } $\renv_Q - \renv = \emptyset$ \textbf{ and } $\renv_S - \renv = \emptyset$}
        \State \rterm{L} $\gets$ \rterm{Q'} $+$ \rterm{S'}  \Comment{Both \rterm{Q'} and \rterm{S'} have the same environment and are multiplicities,}
        \State \Return \rterm{L} \Comment{so we can merge the \rexprs (add the multiplicities)}
      \EndIf
      \State \Return ($\renv_Q -  \renv$) \rterm{*} \rterm{Q'} \rterm{+} ($\renv_S - \renv$) \rterm{*} \rterm{S'} \label{alg:stash-end}%
    \ElsIf{\rterm{R} \textbf{ matches } \rterm{proj(X, Q)}}
      \LinesComment{Omitted: Rewrites from \cref{sec:proj} are applied first, otherwise pass through to the nested \rexpr below.}
      \State $\cdots$
      \LinesComment{Note: Since variable are uniqified, the system does not have to worry about clobbering the slot of another variable with the same name in the surface syntax.}
      \State \rterm{Q'} $\gets$ \Call{\SimplifyFast}{\protect\rterm{Q}, \renv}
      \State \rterm{Q'} $\gets$ \rterm{(X=}{\renv}\rterm{[X])*Q'}  \Comment{Save the value of the variable X into the \rexpr}
      \State \Return \rterm{proj(X, Q')}
    \ElsIf{\rterm{R} \textbf{ matches } \rterm{(A=sum(X, Q))}} \Comment{Match aggregators, exemplified here by \rterm{sum}}
      \LinesComment{Omitted: The rules described in \cref{sec:aggregation_rewrites} for aggregation are run first.}
      \State $\cdots$
      \LinesComment{In the case that none of those rewrites match, the body of the aggregator is matched similar to the projection case.}
      \State \rterm{Q'} $\gets$ \Call{\SimplifyFast}{\protect\rterm{Q}, \renv}
      \State \rterm{Q'} $\gets$ \rterm{(X=}{\renv}\rterm{[X])*Q'}
      \State \Return \rterm{(A=sum(X, Q'))}
    \ElsIf{\rterm{R} \textbf{ matches } \rterm{(A=count(Q))}}
      \LineComment{Omitted: The rules described in \cref{sec:aggregation_rewrites} for the count expression}
      \State $\cdots$
      \State \rterm{Q'} $\gets$ \Call{\SimplifyFast}{\protect\rterm{Q}, \renv} \Comment{If no rules match, simplify the inner expression}
      \If{\rterm{Q'} $\in$ $\theMults$}  \Comment{\rterm{Q'} is a multiplicity, assign its value to \rterm{A}}
        \State \Return \rterm{(A=Q')} 
      \EndIf
      \State \Return \rterm{(A=count(Q'))}    
    \ElsIf{\rterm{R} \textbf{ matches } $\rterm{is(f(T}_1, \ldots \rterm{, T}_n\rterm{)}^m\rterm{,F)}$} \Comment{\cref{rr:is_call}}
      \LinesComment{We prevent unbounded amounts of recursive inlining.  We track the number of times a rule has be applied.  If the recursion limit is exceeded, we simply refuse to rewrite further.}
      \If{$m > \textsc{recursionLimit}$}
          \Return \rterm{R}
      \EndIf
      \LinesComment{The code below, implements a single step of inlining.
        We lookup a \rexpr by name corresponding to a user-defined rule.
        We replace placeholder variables $\rterm{X}_i$ with the variables at this call site, and uniqify variables within the expression.
        }
      \State \rterm{R'} $\gets$ \Call{LookupNamed}{\protect\rterm{f}}
      \State \rterm{R'} $\gets$ \rterm{R'}$\{\rterm{X}_1 \mapsto \rterm{T}_1, \cdots, \rterm{X}_n \mapsto \rterm{T}_n\}$
      \State \rterm{R'} $\gets$ \Call{UniqifyVars}{\protect\rterm{R'}}
      \State \rterm{R'} $\gets$ \rterm{R'}$\{f(\cdots)^0 \to f(\cdots)^{m+1}\}$ \Comment{All calls are annotated with the depth $m+1$}
    \State \Return \rterm{R'}

    \ElsIf{\rterm{R} \textbf{ matches } \rterm{plus(I,J,K)}}
      \LinesComment{The rules for \emph{plus} are equivalent to \cref{rr:plus1,rr:plus2,rr:plus3,rr:plus4} given in \cref{sec:built-ins}}
      \If{\rterm{I} $\in$ \renv \textbf{ and } \rterm{J} $\in$ \renv}
        \State \Return \rterm{(K=}{\renv}[\rterm{I}{}] + {\renv}[\rterm{J}{}]\rterm{)}  \Comment{Compute the addition and set \rterm{K} to the result.}
      \ElsIf{\rterm{I} $\in$ \renv \textbf{ and } \rterm{K} $\in$ \renv}
        \State $\cdots$ \Comment{Other modes similar to the above, omitted for brevity}
      \Else
        \State \Return \rterm{R} \Comment{Return unchanged in the case that it can not run}
      \EndIf
    \ElsIf{\rterm{R} \textbf{ matches } $\cdots$}
      \LinesComment{Omitted: Other matching rules are very similar to the rewrites presented in the main paper.}
      \State $\cdots$
    \Else
      \State \Return \rterm{R} \Comment{In the case that no rewrite matches, then the original \rexpr is returned}
    \EndIf
  \EndFunction
\end{algorithmic}

The main purposes of \textsc{\SimplifyFast} are to handle simple rewrites
including evaluating built-in operations, expanding aggregations and function
calls by one step, and carrying out equality propagation
(\cref{sec:equality_prop}).  If we directly used the rewrite system of the main
paper, then we would have to deliberately apply commutative, associative, and
distributive rewrites to obtain a subterm of the form to which the equality
propagation rewrite applies.

Notice above that equality constraints are \emph{absorbed} into the environment.
When one of these equalities is dropped from the environment (line
\ref{alg:absorb-into-env}), it is restored as a conjunct at the front of the
\rexpr (lines \ref{alg:stash-start} and \ref{alg:stash-end}).  Since
\Call{\SimplifyFast}{} traverses the \rexpr from left to right, this
rearrangement ensures that the next time it is called, the environment will get
fully populated with these equality constraints before \Call{\SimplifyFast}{}
moves rightward to rewrite other constraints in the expression that might need
to look up the values of variables in the environment.

For \Call{\SimplifyComplete}{}, the environment may also contain other types of
constraints, which allows for the discovery of more opportunities to apply
rewrite rules.  In addition, \Call{\SimplifyComplete}{} uses a more aggressive
technique to ensure that constraints that participate in the same rewrite rule
can find one another.  Because the order in which constraints are represented
inside of an \rexpr does not influence the \rexpr's semantic interpretation
(\cref{rr:commutativity_times}), this rewriting pass operates using two passes.
The first pass scans the conjunctive part of the \rexpr using
\Call{\findConj}{} to collect the
relevant constraints into the environment \renv.  This pass does not perform any
rewriting as it does not necessarily have a complete perspective on the
\rexpr.  Once all of the conjunctive constraints are collected, a second pass
using \Call{\SimplifyComplete}{} can perform efficient \emph{local}
rewrites on the \rexpr, while also consider the environment \renv.  This
combination allows for higher-level constraints to be combined, similar to how built-in
values are propagated.  For example, two \rterm{lessthan} constraints can be combined as in
\cref{rr:lessthan_combine} to infer a third \rterm{lessthan} constraint.

\begin{algorithmic}[1]
  \Function{\SimplifyComplete}{\protect\rterm{R}, \renv}
    \LinesComment{Rewrites which require a larger context are included in this routine.}
    \If{\rterm{R} \textbf{ matches } \rterm{Q+S}}
      \State $\renv_Q$ $\gets$ \Call{Copy}{\renv}
      \State $\renv_S$ $\gets$ \Call{Copy}{\renv}
      \State \rterm{Q'} $\gets$ \Call{\SimplifyCompletePartition}{\protect\rterm{Q}, $\renv_Q$} %
      \State \rterm{S'} $\gets$ \Call{\SimplifyCompletePartition}{\protect\rterm{S}, $\renv_S$}
      \State \renv $\cup$= $\renv_Q \cap \renv_S$  \Comment{Note: \renv may now contain constraints as well as bindings to variables}
      \State \Return $(\renv_Q - \renv)$ \rterm{*Q' +} $(\renv_S - \renv)$ \rterm{*S'} \Comment{Constraints not in \renv are added back to their sub-\rexprs
      }
    \ElsIf{\rterm{R}\textbf{ matches }\rterm{Q*S}}
      \State \rterm{Q'} $\gets$ \Call{\SimplifyComplete}{\protect\rterm{Q}, \renv} \Comment{Similar to the conjunctive rule for \textsc{\SimplifyFast}, but recursive with \textsc{\SimplifyComplete}}
      \If{\rterm{Q'} \textbf{ matches } \rterm{0}}
        \Return \rterm{0}  \Comment{The first conjunct has failed, so it can stop processing early. Implements \cref{rr:zero_mult}}
      \EndIf
      \State \rterm{S'} $\gets$ \Call{\SimplifyComplete}{\protect\rterm{S}, \renv}
      \If{\rterm{S'} \textbf{ matches } \rterm{0}}
        \Return \rterm{0} \Comment{Implements \cref{rr:zero_mult}}
      \EndIf
      \If{\rterm{Q'} \textbf{ matches } \rterm{1}}
        \Return \rterm{S'} \Comment{Implements \cref{rr:one_mult}}
      \EndIf
      \If{\rterm{S'} \textbf{ matches } \rterm{1}}
        \Return \rterm{Q'} \Comment{Implements \cref{rr:one_mult}}
      \EndIf
    \ElsIf{\rterm{R}\textbf{ matches }\rterm{lessthan(A, B)}} \Comment{Example of constraint-joining rules \cref{sec:built-ins}}
      
      \If{\rterm{lessthan(B, A)} $\in$ \renv}
        \State \Return \rterm{0}  \Comment{In this case, there are two constraints that are inconsistent, so it rewrites as \rterm{0}.}
      \ElsIf{\rterm{lessthan(B, C)} $\in$ \renv \textbf{ and } \rterm{lessthan(A, C)} $\not\in$ \renv}%
        \LinesComment{This example case implements the transitivity rule for \rterm{lessthan} (described in \cref{sec:built-ins}).  Below, we add \hspace{5mm} \linebreak the new conjunct to the environment (optional) and the \rexpr returned.}
        \State \renv $\cup$= \rterm{lessthan(A, C)}
        \State \Return \rterm{lessthan(A, B)*lessthan(A, C)} \label{alg:add-lessthan-constraint}
      \ElsIf{\rterm{lessthan(C, A)} $\in$ \renv \textbf{ and } \rterm{lessthan(C, B)} $\not\in$ \renv}
        \State $\ldots$ \Comment{Similar to the above for the other side of propagating with lessthan}
      \EndIf

      \State \Return \rterm{R}
    \ElsIf{\rterm{R} \textbf{ matches } $\rterm{(X=}f\rterm{(U}_1\rterm{,}\ldots\rterm{,U}_n\rterm{))}$} \Comment{Handle unification of non-ground terms}
      \If{$\rterm{(X=}g\rterm{(B}_1\rterm{,}\ldots\rterm{,B}_m\rterm{))}$ $\in$ \renv \textbf{ and } ($g \neq f$ \textbf{ or } $n \neq m$)}  \Comment{Mismatched functors, \cref{rr:struct_unify1}}
        \State \Return \rterm{0} \Comment{These functors are incompatible}
      \EndIf
      \If{$\rterm{(X=}f\rterm{(V}_1\rterm{,} \ldots\rterm{,V}_n\rterm{))}$ $\in$ \renv} \Comment{Matched functor, \cref{rr:struct_unify2}}
        \State \Return $\rterm{(U}_1\rterm{=V}_1\rterm{)*}\cdots\rterm{*(A}_n\rterm{=B}_n\rterm{)}$  \Comment{Unify the nested variables}
      \EndIf

      \If{ \Call{OccursCheck}{\protect\rterm{X}, $\protect\rterm{U}_1$, $\ldots$, $\protect\rterm{U}_n$} } \Comment{Occurs to identify cycles like: \rterm{X=f(Z), Z=f(X)} }
        \State \Return \rterm{0} \Comment{for which there does not exist a solution, hence return \rterm{0}, \cref{rr:occurs_check}}
      \EndIf
      \State \Return \rterm{R}
    \Else
      \LinesComment{Omitted: We do not show pattern-matching branches that are the same as \textsc{\SimplifyFast} above. In the omitted branches, the recursive calls are made to \Call{\SimplifyComplete}{} instead of \textsc{\SimplifyFast}.}
      \State $\cdots$
    \EndIf
  \EndFunction

  \Function{\SimplifyCompletePartition}{\protect\rterm{R}, \renv}
    \LinesComment{This routine is called on the top-level expression, and on each disjunct of a disjunctive sub-expression.}
    \LinesComment{First pass: traverse R, adding all conjunctive constraints to the environment}
    \State \renv $\cup$= \Call{\findConj}{\protect\rterm{R}}
    \LinesComment{Second pass: use the environment to simplify this branch}
    \State \rterm{R'} $\gets$ \Call{\SimplifyComplete}{\protect\rterm{R}, \renv}
    \For{\rterm{Q+S} $\in$ \Call{\findDisj}{\protect\rterm{R'}}} \Comment{Branches may rewrite further}
      \For{\rterm{B} $\in$ [\rterm{Q,S}{}]} \Comment{Loop over the branches of the disjunction}
        \State $\rterm{R}_{B}$ $\gets$ $\rterm{R'}\{\rterm{(Q+S)} \mapsto \rterm{B}\}$  \Comment{Replace the disjunction with only one of its branches}
        \State $\renv_B$ $\gets$ \Call{Copy}{\renv}
        \State $\rterm{R}_{B}\rterm{'}$ $\gets$ \Call{\SimplifyComplete}{$\protect\rterm{R}_B$, $\renv_B$}
        \If{$\rterm{R}_{B}\rterm{'}$ \textbf{ matches } \rterm{0}}
          \State \rterm{R'} $\gets$ $\rterm{R'}\{\rterm{B} \mapsto \rterm{0}\}$ \Comment{B was fully eliminated}
        \Else
          \State \rterm{R'} $\gets$ $\rterm{R'}\{\rterm{B} \mapsto ({\renv}_B - \renv) \ \, \rterm{ * B}\}$  \Comment{New constraints are added to this branch. }
        \EndIf
      \EndFor
    \EndFor
    \State \Return \rterm{R'}
  \EndFunction

  \Function{\findConj}{\protect\rterm{R}}
    \LinesComment{Returns an environment containing all of the conjunctive expressions in R.}
    \If{\rterm{R} \textbf{ matches } \rterm{Q+S}}
      \Return $\emptyset$ \Comment{Do not recurse into + expressions}
    \EndIf
    \If{\rterm{R} \textbf{ matches } \rterm{Q*S}}
      \Return \Call{\findConj}{\protect\rterm{Q}} $\cup$
      \Call{\findConj}{\protect\rterm{S}}
    \EndIf
    \LinesComment{Recurse on subexpression of aggregation and projection, merge their environments}%
    \State \renv $\gets \{$\rterm{R}{}$\}$
    \For{\rterm{S} $\in$ \Call{SubExpressions}{\protect\rterm{R}}}
      \State \renv $\cup$= \Call{\findConj}{\protect\rterm{S}}
    \EndFor
    \State \Return \renv
  \EndFunction

  \Function{\findDisj}{\protect\rterm{R}}

    \LinesComment{Returns a list of disjunctive expressions.  For example, in the case that this is called on the \rexpr \rterm{(R0+R1)*(R2+R3)}, then it will return the list of \rexprs: \texttt{[\rterm{R0+R1}, \rterm{R2+R3}{}]}.  This allows for \Call{\SimplifyComplete}{} to identify places in the \rexpr where it can branch by selecting \rterm{R0} or \rterm{R1} for example.}
    \If{\rterm{R} \textbf{ matches } \rterm{Q+S}}
      \State \Return [\rterm{R}{}] \Comment{This only returns the \emph{conjunctive} constraints, so it does not recurse on Q or S}
    \EndIf
    \LinesComment{Recurse on subexpression of aggregation and projection, merge their branches}
    \State b $\gets$ [ ]
    \For{\rterm{S} $\in$ \Call{SubExpressions}{\protect\rterm{R}}}
      \State b.concatenate(\Call{\findDisj}{\protect\rterm{S}}) \Comment{This recurses into conjunctive expressions, therefore combining all disjunctions that it finds into a single list}
    \EndFor
    \State \Return b
    \EndFunction

\end{algorithmic}

%
%
%
%
%
%
%
%

%

\end{document}